\documentclass{jfm}
\usepackage{graphicx}
\usepackage{newtxtext}
\usepackage{newtxmath}
\usepackage{natbib}
\usepackage{hyperref}
\hypersetup{
    colorlinks = true,
    urlcolor   = blue,
    citecolor  = black,
    linkcolor = black
}
\usepackage[dvipsnames]{xcolor}
\usepackage[cal=cm]{mathalpha}
\usepackage{mathrsfs}
\usepackage{soul}

\newcommand{\RomanNumeralCaps}[1]

\definecolor{cream}{RGB}{232,227,211}
\definecolor{revisions}{RGB}{0,0,0}
\definecolor{revisions2}{RGB}{0,0,0}
\definecolor{revisions3}{RGB}{0,0,0}
\definecolor{revisions4}{RGB}{0,0,0}
\definecolor{revisions5}{RGB}{0,0,0}
\definecolor{revisions6}{RGB}{0,0,0}
\definecolor{ref1}{RGB}{0,0,0}
\definecolor{ref2}{RGB}{0,0,0}
\definecolor{ref3}{RGB}{0,0,0}

\definecolor{cream}{RGB}{222,217,201}
\definecolor{graytable}{RGB}{200,200,200}

\newcommand{\ovc}{\overrightarrow}
\newcommand{\sigs}{\sigma^{\prime \star}}
\newcommand{\DPS}{\Delta P^{\star}}
\newcommand{\sigp}{\sigma_{zz}^\prime}
\newcommand{\us}{\tilde{u}_s}

\newcommand{\sigf}{\sigma^\prime_\F}
\newcommand{\zt}{\tilde{z}}

\newcommand{\mK}{\mu K}
\newcommand{\ez}{\overrightarrow{e_z}}

\newcommand{\F}{\mathcal{F}}

\title{Friction modifies the quasistatic mechanical response of a confined, poroelastic medium}
\author{T. Desclaux \aff{1,2,3}, C. Cuttle\aff{3}, C.~W.~MacMinn \aff{3} and O. Liot\aff{1,2} \corresp{\email{olivier.liot@imft.fr}}}
\affiliation{\aff{1}Institut de Mécanique des Fluides de Toulouse, Université de Toulouse, CNRS, Toulouse, France \\
\aff{2}LAAS-CNRS, Université de Toulouse, CNRS, Toulouse, France \\
\aff{3}Department of Engineering Science, University of Oxford, Oxford, OX1 3PJ, United Kingdom}

\begin{document}
\maketitle

\begin{abstract}
The mechanical response of elastic porous media confined within rigid geometries is central to a wide range of industrial, geological, and biomedical systems. However, current models for these problems typically overlook the role of wall friction, and particularly its interaction with confinement. Here, we develop a theoretical framework to describe the interplay between the mechanics of the medium and Coulomb friction at the confining walls for slow, quasistatic deformations in response to two canonical uniaxial forcings: piston-driven loading {\color{ref2}(\textit{i.e.}, an imposed effective stress at the top boundary)} and fluid-driven loading  {\color{ref2}(\textit{i.e.}, an imposed fluid pressure at the top boundary)} followed by unloading. We find that, during compression, the stress field evolves according to a quasistatic advection-diffusion equation, extending classical poroelasticity results. The magnitude of friction is controlled by a single dimensionless number ($\F$) proportional to the friction coefficient and the aspect ratio of the confining geometry. During decompression, a portion of the solid matrix remains stuck due to friction, leading to hysteresis and to the propagation of a slip front. In piston-driven loading, the frictional stress is directly coupled to the solid effective stress, leading to exponential damping of the loading and striking changes to the displacement field. However, this coupling limits the energy dissipated by friction. In fluid-driven loading, the pressure gradient locally adds energy, {\color{ref3}decoupling elastic energy storage and frictional energy dissipation}. The displacement remains qualitatively unchanged but is quantitatively reduced due to large energy dissipation. In both cases, friction can have a substantial impact on the apparent mechanical properties of the medium.

\end{abstract}

\begin{keywords}
Porous media, wet granular media, dry granular media (the list of available keywords is at \href{https://www.cambridge.org/core/journals/journal-of-fluid-mechanics/information/list-of-keywords}{https://www.cambridge.org/core/journals/journal-of-fluid-mechanics/information/list-of-keywords}
\end{keywords}

\section{Introduction}
In a variety of applications, a deformable porous medium can be placed inside a confining container made of rigid and impermeable material. For example, this setup is common in filtration experiments where filter cakes are housed in compression-permeability cells \citep{ruth_studies_1935}. This scenario is also encountered in geo-engineering, where soil samples are consolidated in consolidation cells, and in medicine, particularly when a tumour develops within a confined environment \citep{delarue_compressive_2014}. These porous media are then generally in contact with the confining structure, permitting the exertion of both normal and tangential forces between the two bodies.

{\color{revisions5} The interaction between a porous medium and a confining structure has been widely studied, beginning with dry granular media where the permeating fluid (air) can be neglected. A key classical result is the Janssen effect \citep{janssen_versuche_1895, sperl_experiments_2006}, which arises in tall granular columns: due to interparticle interactions, vertical stresses are redirected laterally toward the container walls, where they are partially supported by friction. As a result, the vertical stress saturates exponentially with depth over a characteristic length $\lambda_\F = R/(2\mu K)$, where $R$ is the column radius, $\mu$
is the coefficient of wall-grain friction, and $K$ is the stress redirection coefficient, a dimensionless constant of order one analogous to the Poisson ratio for elastic solids. This modelling has been extended to liquid-saturated porous media, omitting the viscous pressure gradient induced by fluid-flow \citep{taylor_fundamentals_1948, aguilar-gonzalez_janssen_2025}. Following this methodology, geotechnical investigations revealed that measured soil properties depend on the sample's height $L$ relative to its radius $R$ \citep{taylor_fundamentals_1948,lovisa_tall_2015}, leading to the standard practice in soil mechanics of limiting the ratio $L/R$ to 0.8 \citep{astm_international_standard_2004}. In filtration applications, \citet{lu_stress_1998} incorporated wall friction through continuum modelling to refine predictions from compression-permeability cells, demonstrating that incorporating friction in the continuum model reduced the deviation between predictions and experiment by a factor of two. However, these studies typically simplify the interactions between frictional effects and fluid-solid interactions, either by ignoring the impact of viscous pressure gradients on the solid stresses \citep[\textit{e.g.}][]{lu_stress_1998} or by neglecting the impact of friction on the consolidation process \citep[\textit{e.g.}][]{lovisa_tall_2015}. Moreover, they remain limited to monotonic loading and provide no insights into unloading or cyclic behaviour.

A second approach has prioritised fluid-solid coupling while neglecting wall friction.} For example, \citet{beavers_flow_1975} documents an experiment where a polyurethane sponge is confined in a square channel and compressed by either a fluid flow {\color{ref2}or an external mechanical load, acting in the same way as a permeable piston}. These authors observed poor agreement between experiment and modelling, which they attributed in part to the absence of friction in their model. Later, \citet{parker_steady_1987} and \citet{lanir_nonlinear_1990} conducted experiments using a sponge sample with a slightly reduced width compared to the dimensions of their experimental channel, thereby strategically mitigating the influence of wall friction. However, the resulting gap between the sponge and the walls allowed the flow to partially bypass the sponge, such that they could not simultaneously fit the volume flux and the deformation using a uniaxial  theoretical model. More recently, \citet{hewitt_flow-induced_2016} conducted similar experiments on a packing of hydrogel beads. The authors derived a model to explain their experimental results, neglecting wall friction. The predictions agree with the compressive phase of their experiments, during which the applied pressure head is monotonically increased. However, the authors report unexpected hysteresis during decompression, both in the flow rate and macroscopic strain observed in their fluid-driven experiments and in mechanical stress-strain tests conducted separately. The authors associated this hysteresis with granular rearrangements rather than wall friction, in part because hydrogel beads are known to be particularly slippery \citep{cuccia_pore-size_2020}. This interpretation is consistent with previous studies on granular media, which reported extensive rearrangements in hydrogel packings \citep{macminn_fluid-driven_2015}. However, as explored in more detail below, wall friction can cause qualitatively similar hysteretic behaviour.

Two notable exceptions that bridge these two approaches are the work of Lutz \citep{lutz_method_2021,lutz_frictional_2021} and Li \citep{li_fluid-filled_2022}. Lutz conducted experiments with a stack of latex foam discs in a cylindrical pipe. The foam discs were cut to match the cylinder's inner diameter, and the fluid pressure was measured at multiple locations within the porous medium. They conducted piston- and fluid-driven compression-decompression cycles and observed hysteresis in the displacement-pressure curves, which they attributed to wall friction. This wall friction was observed in both loading scenarios, and its relative importance was found to depend on the system's aspect ratio. Importantly, they identified striking differences between piston-driven and fluid-driven scenarios: during fluid-driven cycles, the foam remained stuck throughout much of the decompression phase, whereas it slipped readily during piston-driven decompression. They developed a discrete model incorporating friction, which qualitatively reproduced their experimental observations. However, the absence of a continuum framework, combined with uncertainty in the size of their disks,  limited the comparison between loading scenarios and their understanding of the fundamental poroelastic behaviour. Along similar lines, \citet{li_fluid-filled_2022} used a novel photoporomechanical technique to visualise the effective-stress field during classical poroelastic consolidation, highlighting the qualitative impact of wall friction and proposing a continuum model with a simplified friction term.

\begin{figure}
\begin{center}
\includegraphics[width=9cm]{"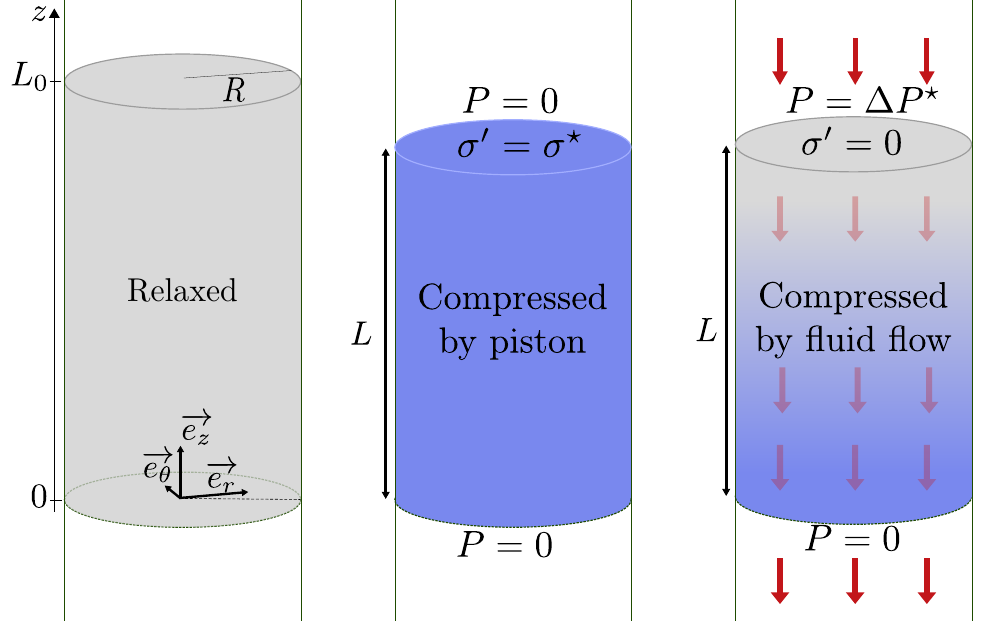"}  
\end{center}
\caption{
A confined cylindrical porous medium is initially at rest (left) and is then compressed either by a permeable piston (centre) or a fluid flow (right). The shading illustrates the level of stress experienced by the solid matrix in the absence of friction, in the steady state: the stress level is uniform in piston-driven compression, and increases linearly from top to bottom in the fluid-driven compression (see section \ref{subsub_poroelast}).}
\label{FGR_PresentationAxes}
\end{figure}

In summary, existing studies address two canonical scenarios, piston-driven and fluid-driven compression of a confined porous medium (figure \ref{FGR_PresentationAxes}), but fail to capture the full three-way coupling between solid deformation, interstitial fluid flow, and wall friction. In piston-driven scenarios, cyclic loading remains undocumented, and the mechanisms governing hysteresis and energy dissipation are poorly understood. In fluid-driven scenarios, no continuum framework exists to represent the frictional poroelastic response. Here, we study this {\color{revisions6}three-way} coupling theoretically. To do so, we develop a uniaxial poroelastic continuum model that accounts for Coulomb-like wall friction driven by stress redirection. {\color{ref3}We then focus our analysis on the quasistatic case, where the movements of the solid matrix are so slow that poroelastic transients can be neglected}. This allows us to derive analytical expressions for the effective stress, the solid displacement, and the energy dissipation during compression and decompression of a medium during loading by either a piston or a fluid flow. In the following, section $\ref{sec_Modelling}$ presents the general modelling framework and section \ref{subsec_analyticalsolution} presents the derivation of analytical solutions for quasi-static loading under a few simplifying assumptions.

\section{Modelling}
\label{sec_Modelling}
\subsection{Mechanical behaviour}
\subsubsection{Governing equations}
This article examines the interaction between friction and linear poroelasticity, leaving certain complexities of large deformation poromechanics for future work. In this section, the key governing equations are briefly presented to describe the coupling of solid matrix deformation and fluid flow. The interested reader can find additional details in \citet{coussy_poromechanics_2004} and \citet{macminn_large_2016}.

We assume that fluid flows through the medium according to Darcy's law with an isotropic permeability $k$:
\begin{equation}
\phi (\overrightarrow{v_f}  - \overrightarrow{v_s}) = -\dfrac{k}{\eta} \overrightarrow{\nabla} P , 
\label{eqn_Darcy}
\end{equation}
where $\phi$ is the porosity, $\overrightarrow{v_f}$ is the velocity of the fluid (averaged over the fluid phase),  $\overrightarrow{v_s}$ the solid velocity, $\eta$ the fluid dynamic viscosity and $P$ the fluid pressure.

Taking the fluid phase to be incompressible, mass conservation  gives
\begin{equation}
\partial_t \phi + \overrightarrow{\nabla} \cdot (\phi \overrightarrow{v_f})=0,
\label{eqn_consmass_f}
\end{equation}
where $\partial_t$ is the partial derivative with respect to time.

Similarly, the solid phase is also assumed to be incompressible and mass conservation for the solid phase gives
\begin{equation}
-\partial_t \phi + \overrightarrow{\nabla} \cdot \left( (1-\phi) \overrightarrow{v_s} \right)=0.
\label{eqn_consmass_s}
\end{equation}

One can then define the total flux $\overrightarrow{q}$ as
\begin{equation}
\overrightarrow{q} = \phi \overrightarrow{v_f} + (1-\phi) \overrightarrow{v_s},
\label{eqn_q}
\end{equation}
and, from equations \eqref{eqn_consmass_f} and \eqref{eqn_consmass_s}, it follows that
\begin{equation}
\overrightarrow{\nabla} \cdot \overrightarrow{q} = 0.
\label{eqn_qconst}
\end{equation}

The Terzaghi decomposition of stress is adopted \citep{terzaghi_theoretical_1943}, in which the total stress $\boldsymbol{\sigma}$ is equal to
\begin{equation}
\boldsymbol{\sigma} = \boldsymbol{\sigma'} - P\mathsfbi{I},
\label{eqn_Terz}
\end{equation}
with $\boldsymbol{\sigma'}$ the Terzaghi effective stress (\textit{i.e.}, the component of the stress that deforms the solid), and $\mathsfbi{I}$ the identity tensor, where the tension-positive sign convention from solid mechanics is adopted \citep{francois_mechanical_2012}.

In the absence of inertia and gravity, mechanical equilibrium is given by
\begin{equation}
\overrightarrow{\nabla} \cdot \boldsymbol{\sigma} = \overrightarrow{0}
\end{equation}
which, together with equation \eqref{eqn_Terz}, implies
\begin{equation}
\overrightarrow{\nabla} \cdot \boldsymbol{\sigma^\prime} = \overrightarrow{\nabla} P .
\label{eqn_Terzaghi}
\end{equation}

{\color{revisions5}For simplicity, we now assume that the deformations of the solid matrix are small. However, we expect the physical reasoning and qualitative observations below to generalise readily to large deformations. In detail, if the solid matrix experiences a typical displacement of $u_s$, then the characteristic scale of the strain is $u_s/L_0$, with $L_0$ the typical relaxed length of the porous medium, and we assume that $\left| u_s/L_0 \right| \ll 1$. This assumption has four important consequences. {\color{ref1}First, the permeability is assumed to be constant and uniform. Second,} the strain tensor $\boldsymbol{\varepsilon}$ is related to the solid displacement $\ovc{u_s}$ via
\begin{equation}
\boldsymbol{\varepsilon} = \dfrac{1}{2} \left( \boldsymbol{\nabla} \overrightarrow{u_s} + (\boldsymbol{\nabla}  \overrightarrow{u_s} )^T \right),
\label{straindef}
\end{equation}
and the stress-strain relationship follows linear isotropic elasticity \citep[\textit{e.g.}][]{francois_mechanical_2012},
\begin{equation}
\boldsymbol{\sigma^\prime} = (\mathcal{M}-\lambda) \boldsymbol{\varepsilon}  + \lambda \text{Tr} (\boldsymbol{\varepsilon}) \mathsfbi{I},
\label{eqn_ElasticityTensor}
\end{equation}
where $\mathcal{M}$ and $\lambda$ are the oedometric modulus and the first Lamé coefficient, respectively.
 Third, the porosity is related to the solid displacement  via \citep[e.g][]{macminn_large_2016} 
\begin{equation}
\label{eqn_StrainDef}
\ovc\nabla \cdot \ovc{u_s} =  \dfrac{\phi-\phi_0}{1-\phi_0},
\end{equation}
with $\phi_0$ the uniform relaxed porosity. Fourth, the solid velocity is related to the solid displacement via $v_s = \partial{u_s}/\partial{t}$, such that 
\begin{equation}
    \ovc{\nabla} \cdot \ovc{v_s} = \ovc{\nabla} \cdot (\partial_t\ovc{u_s}) = \partial_t \left( \ovc{\nabla} \cdot \ovc{u_s} \right).
    \label{eqn_divergencevs}
\end{equation}

Combining equations \eqref{eqn_Darcy} and \eqref{eqn_q} with equation \eqref{eqn_qconst} gives
\begin{equation}
\ovc{\nabla}\cdot \ovc{v_s} - \ovc{\nabla}\cdot \left(\frac{k}{\eta} \ovc{\nabla} P\right)  = 0.
\end{equation}
Using equations \eqref{eqn_StrainDef} and \eqref{eqn_divergencevs}, the above becomes
\begin{equation}
\frac{\partial_t \phi}{1-\phi_0} - \frac{k}{\eta} \ovc{\nabla} \cdot \left(\ovc{\nabla} P\right)  = 0,
\label{eqn_LinearPoroelastic}
\end{equation}
which is one form of the classical linear poroelastic flow equation. Neglecting frictional effects, equation \eqref{eqn_LinearPoroelastic} can be written in the form of a linear diffusion equation for the porosity field: the fluid pressure is coupled to the effective stress via equation \eqref{eqn_Terzaghi}, and the porosity is coupled to the solid mechanics via equations \eqref{straindef} -- \eqref{eqn_ElasticityTensor}. The above is classical poroelasticity theory {\color{ref3}for incompressible constituents} \citep{coussy_poromechanics_2004}.}

\subsubsection{Boundary conditions, initial values}
\label{subsec_BC}
We consider a porous medium of relaxed length $L_0$ confined within a cylindrical structure of radius $R$ (figure \ref{FGR_PresentationAxes}a), which undergoes uniaxial compression. In the following, we adopt cylindrical coordinates ($\overrightarrow{e_r},\overrightarrow{e_\theta},\overrightarrow{e_z}$), as illustrated in figure \ref{FGR_PresentationAxes}.  As the confining structure is impermeable to both fluid and solid,

\begin{equation}
    \left. \overrightarrow{v_s}\cdot \overrightarrow{e_r} \right|_R =\left. \overrightarrow{v_f}\cdot \overrightarrow{e_r}\right|_R = 0,
\end{equation}
where $ \left. \overrightarrow{v_s}\cdot \overrightarrow{e_r}\right|_R$ and $ \left. \overrightarrow{v_f}\cdot \overrightarrow{e_r}\right|_R$ are, respectively, the solid and fluid velocity normal to the walls.

For downward piston-driven compression, we take the top boundary to be permeable such that the effective stress is equal to the stress applied by the piston and the fluid pressure is null. {\color{ref2}We take the piston to be controlled in loading and, for simplicity, we  take the imposed stress to be uniform over the top boundary \citep[as in][]{macminn_large_2016,hewitt_flow-induced_2016}. Finally, } we take the bottom boundary to be permeable and fixed in place, such that the solid displacement and the fluid pressure are null. Thus, we have 
\begin{equation}
\left\{
    \begin{array}{lll}
 \sigp = -\sigma^{\prime\star}   & \text{and }   P  = 0 & \mathrm{at} ~ z=L  \\
 \ovc{u_s}= \ovc{0} & \text{and }  P = 0 & \mathrm{at} ~ z=0,
\end{array}\right.
\end{equation}
where the location $z=L\approx L_0$ represents the top of the medium, and $z=0$ the bottom of the medium, {\color{revisions5}and where $-\sigma^{\prime\star}$ is the stress applied by the piston, with $\sigma^\star>0$}.

For downward fluid-driven compression, we take the top boundary to be unconstrained, so that the associated mechanical stress is null and the fluid pressure is equal to the imposed operating pressure $\Delta P^\star$. We take the bottom boundary to be again permeable and fixed in place, such that the solid displacement and the fluid pressure are again null. Thus, we have
\begin{equation}
\left\{
    \begin{array}{lll}
    \sigp = 0  & \text{and } P  = \Delta P^\star & \text{at }  z=L \\
    \ovc{u_s} = \ovc{0} & \text{and } P  = 0 & \text{at } z=0,
    \end{array}\right.
\label{eqn_bc_fluid}
\end{equation}

\subsection{Impact of friction}
\subsubsection{Modelling tangential forces and confinement}
In this problem, friction manifests as a nonzero shear stress between the material and the confining walls, $\sigma'_{rz}(r=R) \neq 0$. We next consider the impact of friction on the two problems above.

Because of the axisymmetry, all quantities remain constant along with varying $\theta$. Further, the uniaxial nature of the problem suggests that $u_s$, $v_s$, and $v_f$ will be primarily directed along $\ez$, such that $\ovc{v_f} \approx{} v_f(r,z) \ez,~ \ovc{v_s} \approx{} v_s(r,z) \ez$, and $\ovc{u_s} \approx u_s(r,z) \ez$. From equation \eqref{straindef}, the strain is then 
 
\begin{equation}
\boldsymbol{\varepsilon} = \begin{pmatrix}
0 & 0 & \dfrac{1}{2} \partial_r u_s \\
0 & 0 & 0 \\
\dfrac{1}{2} \partial_r u_s & 0  & \partial_z u_s  
\end{pmatrix}\, ,
\label{DefStrainDepl}
\end{equation}
{\color{revisions5}so that equation \eqref{eqn_StrainDef} implies that
\begin{equation}
    \varepsilon_{zz} =  \dfrac{\phi-\phi_0}{1-\phi_0}.
    \label{eqn_PorosityStrain}
\end{equation}}
From equations \eqref{eqn_ElasticityTensor} and \eqref{DefStrainDepl}, the effective stress  is then
\begin{equation}
\boldsymbol{\sigma^\prime} = \begin{pmatrix}
\lambda \varepsilon_{zz} & 0 & (\mathcal{M}-\lambda) \varepsilon_{rz} \\
0 & \lambda \varepsilon_{zz} & 0 \\
(\mathcal{M}-\lambda) \varepsilon_{rz}  & 0  & \mathcal{M} \varepsilon_{zz}
\end{pmatrix}\, .
\label{eqn_StressStrainRelation}
\end{equation}

Therefore, the $z$-component of equation \eqref{eqn_Terzaghi} becomes
\begin{equation}
\partial_z \sigma^\prime_{zz} + \frac{\partial_r \left(r\sigma^\prime_{rz}\right)}{r}  = \partial_z P.
\label{EQN_EquilibreMec}
\end{equation}
Equation \eqref{EQN_EquilibreMec} can be integrated along $r$ and $\theta$:
\begin{equation}
\int_0^R \int_0^{2\pi} \left(r \partial_z \sigma^\prime_{zz}  + \partial_r (r \sigma^\prime_{rz}) \right) drd\theta = \int_0^R \int_0^{2\pi}   \partial_z P rdrd\theta.
\label{eqn_IntegralZTerzaghi}
\end{equation}
This result is then simplified as follows. First, as mentioned in section \ref{subsec_BC}, we aim to capture mechanical tests and other scenarios that are restricted to compression of the medium, rather than shear mechanics. Therefore, we assume that $\sigma^\prime_{zz}$ is constant over horizontal cross-sections, as is usually done in the classical Janssen modelling:
\begin{equation}
\int_0^R \int_0^{2\pi} r\partial_z \sigma^\prime_{zz} drd\theta =  \pi R^2 \partial_z  \sigma_{zz}^\prime.
\label{eqn_AveNormalStress}
\end{equation}
Second, the $r$-component of Darcy's law (equation \ref{eqn_Darcy}) implies that $\partial_r P = 0$ and, as a result, that 
\begin{equation}
 \int_0^R \int_0^{2\pi} \partial_z P rdrd\theta = \pi R^2 \partial_z P.   
\end{equation}

Third, one may note that
\begin{equation}
\int_0^{2\pi} \int_0^R  \partial_r (r \sigma^\prime_{rz})drd\theta = 2 \pi [r \sigma^\prime_{rz}]_0^R = 2\pi  R \left. \sigma^\prime_{rz} \right|_R,
\end{equation}
where the term $\left. \sigma^\prime_{rz}\right\rvert_R$ is the frictional stress applied by the wall on the solid matrix, which, going forward, we denote as $\sigma^\prime_\F$. 
With the above results, equation \eqref{eqn_IntegralZTerzaghi} reduces to
\begin{equation}
\partial_z  \sigma^\prime_{zz}  + \dfrac{2 {\sigma}^\prime_\F}{R}   =   \partial_zP .
\label{eqn_ApparitionFriction_simple}
\end{equation}
{\color{revisions5}Note that this result was derived assuming uniform $\sigma^\prime_{zz}$ across horizontal sections. More generally, the term $\partial_z \sigma^\prime_{zz}$ can be interpreted as the cross-sectional average defined in equation \eqref{eqn_AveNormalStress}, allowing this formulation to be extended beyond the uniform stress assumption. }

In the following, we explore how a non-zero frictional stress $\sigma^\prime_\F$ modifies the coupling between the solid stress and the fluid pressure. 



\subsubsection{Modelling of friction and stress redirection}
\label{subsec_frictionmodelling}

The simplest way to model wall friction is via Coulomb's law \citep{ludema_friction_1996}. Coulomb's law states that the friction force between two solid bodies in contact opposes the relative motion of the bodies, up to a maximum value proportional to the normal load. The associated constant of proportionality is called the coefficient of friction. Coulomb's law further postulates that this coefficient does not depend on the area of contact or the speed of relative motion. However, it may depend on the existence of relative motion -- a distinction being made between the \textit{static}  coefficient, when the relative speed is null, and the \textit{dynamic} coefficient, when the relative velocity is non-zero. We take the static and dynamic values to be equal for simplicity, and thus write Coulomb's law as
\begin{equation}
\begin{cases}
\sigma_F^\prime =  \mathrm{sgn}(v_s) \mu\, \sigma^\prime_{rr} & \iff \left| v_s \right| > 0, \\

\left| \sigma_F^\prime \right| \leq \mu  \left| \sigma^\prime_{rr}   \right|  &  \iff v_s = 0  ,
\label{eqTangentialStress1}
\end{cases}
\end{equation}
with $\mu$ the (dimensionless) coefficient of friction and $\mathrm{sgn}(v_s)$ the sign of $v_s$. This law is restricted to situations where the normal load is null or \textit{pushing} on the wall ($\sigma'_{rr}(R) \leq 0$). The case where a tangential force is not null in the presence of \textit{pulling} forces ($\sigma'_{rr}  (R) > 0$) corresponds to an adhesive surface and will not be discussed in this study.

The link between $\sigma_{rr}^\prime$ and $\sigma_{zz}^\prime$ can be made in two ways. The first way is through linear elasticity, where equation \eqref{eqn_StressStrainRelation} states that $\sigma^\prime_{rr}~=~[\nu/(1-\nu)]\sigp$, with $\nu$ the Poisson ratio of the solid skeleton. The second way is via the mechanics of granular media, where materials are not simple elastic solids and do not obey linear elasticity. Nonetheless, a classical assumption in granular media is that the vertical stress is redirected in the horizontal direction in a fixed proportion:  $\sigma^\prime_{rr} = K \sigma^\prime_{zz}$, with $K$ a dimensionless constant, sometimes referred to as the Janssen parameter \citep[\textit{e.g.}][]{ovarlez_elastic_2005}. Although not derived from linear elasticity, this relation qualitatively resembles the stress redistribution observed in elastic solids, with $K$ playing the same role as $\frac{\nu}{1-\nu}$, and thus $K\approx 1$ for incompressible materials. Applying the convention from granular materials, equation \eqref{eqTangentialStress1} becomes
\begin{equation}
\begin{cases}
\sigma_F^\prime = \mathrm{sgn}(v_s) \mu K \sigma^\prime_{zz} & \iff  \left| v_s \right| > 0, \\

\left| \sigma_F^\prime \right| \leq \mu K \left| \sigma^\prime_{zz} \right| & \iff v_s = 0 .
\end{cases}
\label{eqn_CoulombIntermediate}
\end{equation}

With equation \eqref{eqn_CoulombIntermediate}, the model is fully specified (\textit{i.e.}, this system of equations is closed). Note that equation \eqref{eqn_CoulombIntermediate} does not determine the value of $\sigma^\prime_\F$ in the case $v_s=0$, but the fact that $v_s=0$ provides sufficient information to determine $\sigma^\prime_\F$ from the other parts of the model -- for example, see equation \eqref{eqn_sigmaF_NOSPEED} in the Appendix. Because many intermediate equations were detailed here, we summarise the equations to be solved in table~\ref{TBL_SystemOfEquations}, comprising a system of eight independent equations in eight unknown variables. 

\begin{table}
  \begin{center}
\def~{\hphantom{0}}
  \begin{tabular}{l|c|c}
Name & Equation & Eqn. nb. \\ \hline 
{\small Darcy }& $\phi  (v_f - v_s) = -\frac{k}{\eta} \partial_z P$ & {\small \ref{eqn_Darcy}} \\ [8pt]  
{\small Total flux $q$} & $q = \phi v_f + (1-\phi) v_s $ & {\small \ref{eqn_q}} \\   [8pt]  
{\small Mass conservation (total)} & $\partial_z q = 0$ & {\small \ref{eqn_qconst}} \\   [6pt]  
{\color{revisions5}{\small Poroelastic flow}} & $\frac{\partial_t \phi}{1-\phi_0} - \frac{k}{\eta} \partial_z^2 P  = 0$ & {\small \ref{eqn_LinearPoroelastic}} \\   [8pt]  
{\small Porosity-strain relationship} & $ \varepsilon_{zz}= \frac{\phi-\phi_0}{1-\phi_0} $ & {\small  \ref{eqn_PorosityStrain}} \\   [9pt]  
{\small Solid constitutive law} & $ \sigma^\prime_{zz} = \mathcal{M} \varepsilon_{zz} $ & {\small  \ref{eqn_StressStrainRelation}} \\   [6pt]  
{\small Mechanical equilibrium} &  $\partial_z \sigma^\prime_{zz} + \frac{2 \sigma^\prime_\F}{R} =  \partial_z P$ & {\small \ref{eqn_ApparitionFriction_simple}} \\   [6pt]  
{\small Coulomb's law} & $ \begin{cases}
\sigma_F^\prime = \mathrm{sgn}(v_s) \mu K \sigma^\prime_{zz} & \iff  \left| v_s \right| > 0, \\[6pt]  
\left| \sigma_F^\prime \right| \leq \mu K \left| \sigma^\prime_{zz} \right| & \iff v_s = 0 .
\end{cases} $ & {\small \ref{eqn_CoulombIntermediate}} \\ 
  \end{tabular}
  \end{center}
  \caption{Summary of the model. Each independent equation is displayed (middle column) with its number for easier reference in text (right column) and a short name (left). The system comprises eight independent equations in 8 unknown variables: $\phi, v_f, v_s, q, P, \varepsilon_{zz}, \sigma_{zz}^\prime$ and $ \sigma_F^\prime$. The other quantities, $k$, $\eta$, $\mathcal{M}$, $\phi_0$, $R$, $\mu$, and $K$, are constant parameters, assumed to be known.}
\label{TBL_SystemOfEquations}
\end{table}

\subsection{Sliding regime}
\label{subsec_assumptions}
The model presented in table~\ref{TBL_SystemOfEquations} is generally intractable to solve analytically. Indeed, $\sigma'_F (z)$ may exhibit discontinuities due to the distinction between stick and slip ($v_s=0$ versus $|v_s|>0$) and its sign depends on the sign of $v_s$. These complications can be substantially simplified by considering, for example, an initially uncompressed medium subjected to a monotonically increasing fluid pressure $\DPS$ or mechanical load $\sigs$. In this scenario, the entire medium undergoes compression ($v_s<0$, for $z\in[0,L]$), and the magnitude of the friction stress everywhere equals its maximum value:
\begin{equation}
\left| \sigma_F^\prime \right| = \mu K \left| \sigp \right| .
\label{eqn_SigmaFricitonSimple}
\end{equation}
This framework extends to arbitrary loading scenarios by analyzing individual sliding regions, defined as domains where $|v_s| > 0$. Within such a region, the  continuity of $v_s$, ensures that it remains uniformly positive or negative. In that case, equation \eqref{eqn_SigmaFricitonSimple} remains valid throughout the entire sliding region, and equation \eqref{eqn_ApparitionFriction_simple} reduces to
\begin{equation}
\partial_z P = \partial_z \sigp +\mathrm{sgn}(v_s) 2 \dfrac{\mu K}{R} \sigp,
\label{eqn_ModelGeneral}
\end{equation}
where $\mathrm{sgn}(v_s)=-1$ when the solid moves downward (\textit{e.g.} during compression), and $\mathrm{sgn}(v_s)=1$ when it moves upward (\textit{e.g.} during decompression).

Applying the porosity-strain relationship (equation \ref{eqn_PorosityStrain}), and linear elasticity (equation \ref{eqn_StressStrainRelation}), to equation \eqref{eqn_LinearPoroelastic} leads to
\begin{equation}
\partial^2_{z} P = \frac{\eta}{k \mathcal{M}}\partial_t \sigp.
\label{dy2Peqdtsig}
\end{equation}
Differentiating equation \eqref{eqn_ModelGeneral} with respect to $z$ and and substituting into equation \eqref{dy2Peqdtsig}  leads to
\begin{equation}
\partial_t \sigp = \frac{k\mathcal{M}}{\eta} \partial^2_z \sigp + \mathrm{sgn}(v_s) \frac{k\mathcal{M}}{\eta}\frac{2 \mu K}{R} \partial_z \sigp.
\label{eqn_AdvDiffStress}
\end{equation}
This linear parabolic partial differential equation describes the evolution of the effective stress during the sliding of a confined porous medium. Equation \eqref{eqn_AdvDiffStress} has the structure of a diffusion-advection equation, with the diffusion term being poroelastic (first term on the right-hand side) and the advection term being frictional (second term on the right-hand side). 

\subsection{Scaling}
\label{subsec_scaling}
In the following, equation \eqref{eqn_AdvDiffStress} and the set of equations presented in table \ref{TBL_SystemOfEquations} are scaled as
\begin{equation}
\begin{aligned}
\tilde{z} = \dfrac{z}{L}, & \quad \tilde{u}_s = \dfrac{u_s}{L}, & \quad  \tilde{t} = \frac{t}{T_\mathrm{pe}}, \\
 \quad \tilde{\sigma}^\prime = \dfrac{\sigp}{\mathcal{M}}, & \quad \tilde{P} = \dfrac{P}{\mathcal{M}},
\end{aligned}
\end{equation}
 where we used tildes to indicate non-dimensional variables, simplified the notation, writing $\tilde{\sigma}^\prime := \tilde{\sigma}^\prime_{zz}$, as this is the only component of the stress tensor that appears in the remainder of the article, {\color{ref3}and where 
\begin{equation}
T_\mathrm{pe}=\frac{\eta L^2}{k\mathcal{M}}
\label{eqn_Tpe}
\end{equation}
is the classical poroelastic time scale \citep{cheng_poroelasticity_2016}.} With these scalings, equation \eqref{eqn_AdvDiffStress} can be written in non-dimensional form as
\begin{equation}
\partial_{\tilde{t}} \tilde{\sigma}^\prime = \partial^2_{\tilde{z}} \tilde{\sigma}^\prime + \mathrm{sgn}(v_s) \frac{2 \mu K L}{R} \partial_{\tilde{z}} \tilde{\sigma}^\prime.
\label{eqn_AdvDiffStressAdim}
\end{equation}

This equation shows that during the sliding of a frictional porous medium, the analogue of the Péclet number is the dimensionless number
\begin{equation}
\F = \dfrac{2\mK L}{R},
\label{eqn_alpha}
\end{equation}
which we name the ``friction number'' \citep[see also][]{li_fluid-filled_2022}.  Equation \eqref{eqn_ModelGeneral} can then be written in non-dimensional form as
\begin{equation}
\partial_{\tilde{z}} \tilde{P} = \partial_{\tilde{z}}  \tilde{\sigma}^\prime
 + \mathrm{sgn}(v_s) \F  \tilde{\sigma}^\prime.
\label{eqn_ModelGeneralAdim}
\end{equation}

This equation shows that $\F$ represents the importance of friction in mechanical equilibrium relative to the other terms of equation \eqref{eqn_ModelGeneralAdim}. For $\F \ll 1$, friction is negligible and the gradients in fluid pressure and effective stress are fully coupled ($\partial_{\tilde{z}} \tilde{P} \approx \partial_{\tilde{z}}  \tilde{\sigma}^\prime$), such that any variation of fluid pressure is transmitted to the solid. However, for $\F \gtrsim 1$, fluid pressure gradients are balanced by a combination of solid stress gradients and the frictional contribution due to $\sigma_F^\prime$. Note that $\F$ is proportional to the coefficient of friction ($\mu$), the coefficient of stress redirection ($K$), and the aspect ratio of the porous medium ($L/R$). Therefore, friction may have significant effects even with a low coefficient of friction, as long as the medium is strongly confined (\textit{i.e.} if the aspect ratio is high). 

For quasi-static loading (\textit{i.e.}, loading that is slow relative to the poroelastic timescale), our problem reduces to a steady advection-diffusion problem. {\color{revisions5}Indeed, in equation \eqref{eqn_AdvDiffStressAdim} the time can be non-dimensionalised by the loading time scale $T_\mathrm{load}$ instead of the poroelastic time scale, which leads to 
\begin{equation}
\frac{T_\mathrm{pe}}{T_{\mathrm{load}}}\partial_{\tilde{t}} \tilde{\sigma}^\prime = \partial^2_{\tilde{z}} \tilde{\sigma}^\prime + \mathrm{sgn}(v_s) \F \partial_{\tilde{z}} \tilde{\sigma}^\prime.
\label{eqn_AdvDiffStressAdim_TpeLoad}
\end{equation} 
so that for sufficiently high $T_\mathrm{load}$, $\frac{T_\mathrm{pe}}{T_{\mathrm{load}}}\partial_{\tilde{t}} \tilde{\sigma}^\prime \ll1$, and the left hand term can be neglected. In this case, integrating equation \eqref{eqn_AdvDiffStressAdim_TpeLoad} with respect to $\zt$ leads to
\begin{equation}
C= \partial_{\tilde{z}} \tilde{\sigma}^\prime + \mathrm{sgn}(v_s) \F \tilde{\sigma}^\prime,
\label{eqn_AdvDiffStressAdimC}
\end{equation} 
where $C$ is a constant that we identify as $\partial_z \tilde{P}$, using equation \eqref{eqn_ModelGeneralAdim}. Therefore, for quasistatic loadings, equation \eqref{eqn_ModelGeneralAdim} can be solved as a differential equation in $\tilde{\sigma}^\prime$ and its spatial derivatives along the $\zt$ coordinate.}

Finally, note that the modelling and results presented here are readily generalised to any geometry with a uniform horizontal cross-section (along the $z$ direction):
\begin{equation} 
\F = \mu KL \frac{\mathcal{P}}{\mathcal{A}},
\end{equation}
where $\mathcal{P}$ represents the perimeter of a cross-section, and $\mathcal{A}$ its area. This ratio reduces to $2/R$ for a cylindrical structure, to $2(H+W)/(HW)$ for a rectangular cross-section of width $W$ and height $H$, and to $2/W$ for slender structures ($H \gg W$).

\renewcommand{\sigp}{\tilde{\sigma}^\prime}
\renewcommand{\sigs}{\tilde{\sigma}^{\prime \star}}
\renewcommand{\DPS}{\Delta \tilde{P}^\star}
\renewcommand{\sigf}{\tilde{\sigma}_F^\prime}

\section{Results}
\label{subsec_analyticalsolution}

\subsection{Results in the absence of friction}
\label{subsub_poroelast}
In the following, we focus on the quasistatic regime and work in terms of dimensionless quantities. In the absence of friction $\F=0$, and equation \eqref{eqn_ModelGeneralAdim} reduces to
\begin{equation}
    \partial_{\zt} \tilde{P} = \partial_{\zt} \tilde{\sigma}^\prime
.\end{equation}

The quasistatic compression driven by a piston (\textit{i.e.}, so-called "drained" compression) is conceptually the simplest case. Our assumptions imply that the fluid has time to exit the porous medium so that viscous pressure gradients are negligible ($\partial_{\zt}\tilde{P}=0$), in which case the problem is simply elastic rather than poroelastic. In this case, the effective stress is uniform and equal to the imposed stress (figure \ref{FGR_PresentationAxes}b):
\begin{equation}
\tilde{\sigma}^\prime = -\sigs,
\label{SANSFRIC_sigmaPiston}
\end{equation}
and the displacement field is linear in $\zt$,
\begin{equation}
\tilde{u_s}= -{\sigs}\zt.
\label{SANSFRIC_usPiston}
\end{equation}

For quasi-static compression by a fluid flow, in contrast, the effective stress increases linearly from the top to the bottom, due to the uniform gradient of fluid pressure (figure \ref{FGR_PresentationAxes}c):
\begin{equation}
\tilde{\sigma}^\prime = - \DPS\left(1 -\zt  \right),
\label{SANSFRIC_sigmaFluid}
\end{equation}
so that the displacement field is quadratic in $\zt$
\begin{equation}
\tilde{u_s}= -\DPS ~ \zt\left( 1- \frac{\zt}{2} \right).
\label{SANSFRIC_usFluid}
\end{equation}

\subsection{Solutions during compression}
\label{subsec_solutioncomp}
We now consider the frictional case under compression. An initially uncompressed medium is subjected to compression by steadily increasing the mechanical forcing. Under these conditions, sliding occurs throughout the entire medium, meaning equation \eqref{eqn_ModelGeneralAdim} is valid everywhere within the domain. 

When compression is driven by a piston, equation \eqref{eqn_ModelGeneralAdim} simplifies to
\begin{equation}
0 = \partial_{\zt} \sigp - {\F} \sigp.
\label{eqn_EqDiffCompPiston}
\end{equation}
In that case, the solution of equation \eqref{eqn_EqDiffCompPiston} is 
\begin{equation}
{\sigp} = -{\sigs} \exp \left[{ -\F \left(1-\zt \right)} \right].
\label{eqn_SigmaPiston}
\end{equation}

The displacement field $\tilde{u}_s$ is then obtained by integrating the strain, directly computed from the stress-strain relationship (equation \ref{eqn_StressStrainRelation}), over the $\zt$ coordinate, and imposing $\us |_{\zt=0}=0$:
\begin{equation}
\us = -\dfrac{\sigs}{ \F} \left\{ \exp \left[{- \F \left(1-\zt \right)}\right]-\exp ({-\F})  \right\} .
\label{eqn_u_sPiston}
\end{equation}

In the case of a fluid-driven compression, the pressure difference across the medium $\DPS$ is imposed, while the effective stress at the top is zero. Therefore, equation~\eqref{eqn_ModelGeneral} may be rewritten by replacing $\partial_{\zt} \tilde{P} = \DPS$ : 

\begin{equation}
\Delta \tilde{P}^\star =  \partial_{\zt} \sigp - \F \sigp .
\label{eqn_EqDiffCompFluid}
\end{equation}
With the boundary condition $\left. \sigp \right|_{\zt=1} = 0$, the solution is
\begin{equation}
\sigp = -\dfrac{\Delta \tilde{P}^\star}{\F} \left\{1- \exp \left[-\F\left(1-\zt \right)\right] \right\}.
\label{eqn_StressFluidComp}
\end{equation}
The displacement field is obtained as before:
\begin{equation}
\tilde{u}_s = -\dfrac{\Delta \tilde{P}^\star}{ \F^2} \left\{  \F\zt+\exp({ -\F}) -\exp \left[ -\F \left(1- \zt \right) \right] \right\}.
\label{eqn_usfluide}
\end{equation}

\begin{figure*}
\begin{center}
\includegraphics[width=\textwidth]{"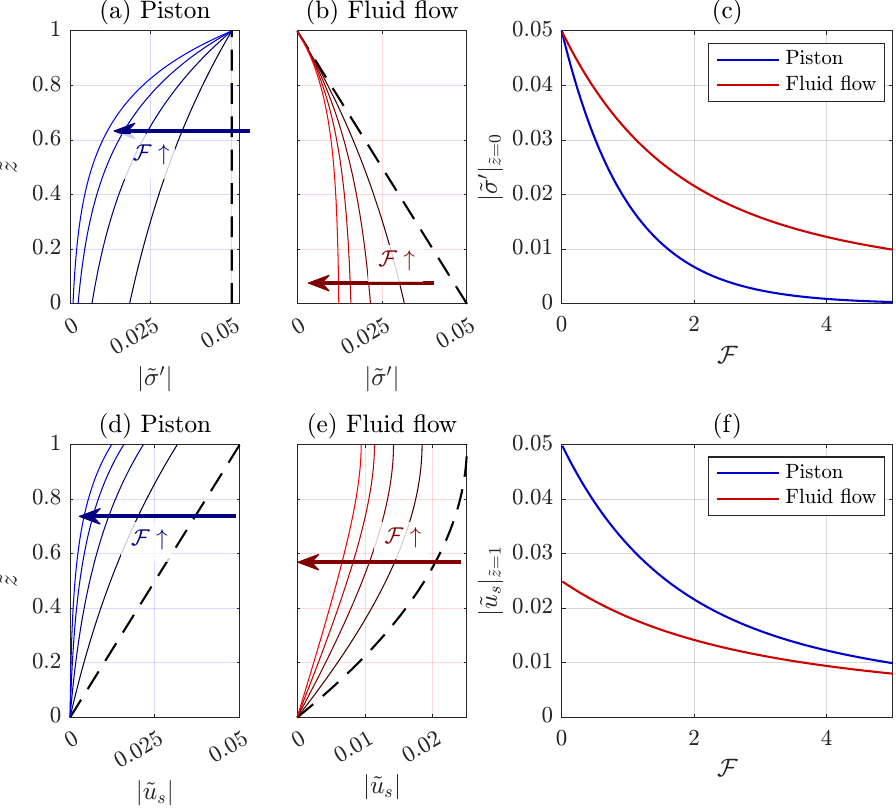"}
\caption{{\color{ref1}Magnitude of the effective stress (top row) and of the relative displacements (bottom row), as a function of relative position ($\zt$), due to forcing by a piston (a and d, equations \ref{eqn_SigmaPiston} and \ref{eqn_u_sPiston}) or by a fluid flow (b and e, equations \ref{eqn_StressFluidComp} and \ref{eqn_usfluide}), with $\F \in \{0,1, 2, 3, 4\}.$ The dotted curve corresponds to the frictionless case ($\F=0$, section\,\ref{subsub_poroelast}), and the coloured arrow points toward increasing friction number. The imposed stress and fluid pressure are fixed at $|\sigs| = \DPS =  0.05$. (c): magnitude of the stress at the bottom of the medium ($\zt=0$) as a function of the friction number (equations \ref{eqn_StressBottom_fluid} and \ref{eqn_StressBottom_piston}). (f): magnitude of displacement of the top of the medium as a function of the friction number (equations \ref{eqn_UsTop_Piston} and \ref{eqn_UsTop_Fluid}).}}
\label{FIG_Comp}
\end{center}
\end{figure*}

{\color{ref1}Equations \eqref{eqn_SigmaPiston}, \eqref{eqn_u_sPiston}, \eqref{eqn_StressFluidComp}, and \eqref{eqn_usfluide} show that $\F$, and either $\sigs$ (for piston-driven loading), and $\DPS$ (for fluid-driven loading) are essential quantities to describe the mechanical response of a confined porous medium subject to wall friction}. The applied stress magnitudes ($\sigs$, $\DPS$) are just scaling factors, whereas the friction number appears in multiple, distinct terms, and thus has a qualitative impact on the solutions.  In figure \ref{FIG_Comp}, the results of equations \eqref{eqn_SigmaPiston} {\color{ref1}to}  \eqref{eqn_usfluide} are plotted for $\F\in \{ 0, 1, 2, 3, 4 \}$.  In the limit of small friction number ($  \F\rightarrow 0$), both the stress field and the displacement field become equivalent to the ones given by equations \eqref{SANSFRIC_sigmaPiston} and \eqref{SANSFRIC_usPiston} in the piston-driven case, and to the ones given by equations \eqref{SANSFRIC_sigmaFluid} and \eqref{SANSFRIC_usFluid} in the flow-driven case. For piston-driven compression, the frictionless stress field is uniform in the medium, so that the stress at the bottom ($\zt=0$) is equal to the imposed stress at the top ($\zt=1$). In the presence of friction, as  $\zt$ decreases, the effective stress decays exponentially over a characteristic dimensionless distance $\mathcal{F}^{-1}$: the friction has dissipated the mechanical energy that was provided on the top of the medium. As a consequence, at high friction ($\F =5$), the bottom of the medium is almost uncompressed, since the stress at the bottom is given by
\begin{equation}
    \sigp|_{\zt=0} = -\sigs \exp(-\F),
    \label{eqn_StressBottom_piston}
\end{equation}
{\color{ref1}which is illustrated in figure \ref{FIG_Comp}\,(c).} Thus, friction cumulatively shields the material from the applied load. This exponential decay of stress and deformation with distance from the loading point is a hallmark of the classical Janssen effect.

For fluid-driven compression, the frictionless stress field varies linearly in the medium, from zero at the top to the imposed hydrostatic pressure difference ($\DPS$) at the bottom. With friction, the stress field deviates exponentially from the frictionless case toward a uniform value over a characteristic dimensionless distance $\F^{-1}$. This uniform value is given by {\color{ref1}the stress at the bottom}:
\begin{equation}
    \sigp|_{\zt=0} = -\DPS \frac{1-\exp(-\F)}{\F},
    \label{eqn_StressBottom_fluid}
\end{equation}
{\color{ref3}which represents the maximum of the stress field magnitude. Indeed, in the fluid-driven case, this stress magnitude remains below $\DPS$ (see figure \ref{FIG_Comp} (b)).}

At high friction ($\F \gg 1$), the stress at the bottom is equivalent to $\DPS/\F  \propto \F^{-1}$. {\color{ref1}Therefore, as it can be observed in figure \ref{FIG_Comp}\,(c), when friction is high, the stress at the bottom is much higher in the fluid-driven compression than in the piston-driven compression}. Friction cannot shield the material far below the top surface from fluid-driven loading because the viscous pressure gradient acts throughout the entire material, not just at the top.

The displacement profiles for both loading scenarios are shown in figure \ref{FIG_Comp}\,(d,e). For piston-driven loading, the frictionless displacement field increases linearly with $\zt$. With friction, the displacement field instead becomes an exponential function of $\zt$, with almost no displacement over an increasingly large part of the medium as friction increases. The displacement of the top of the medium is given by
\begin{equation}
    \left. \us \right|_{\zt=1} = -\sigs \frac{1-\exp(-\F)}{\F},
    \label{eqn_UsTop_Piston}
\end{equation}
{\color{ref1}which is presented in figure \ref{FIG_Comp}\,(f).} For small friction numbers ($\F\ll1$), these displacements are equivalent to $-\sigs(1-\F/2)$, while at high friction numbers ($\F\gg1$), they become equivalent to $-\sigs/\F$. 

For fluid-driven compression, the frictionless displacement field increases quadratically with $\zt$. With friction, the solution retains its key qualitative features, having $\tilde{u}_s$ linear in $\zt$ at $\zt=0$ and $\partial_{\zt} \tilde{u}_s=0$ at $\zt=1$. A quasilinear region emerges in the lower portion of the medium and expands as $\F$ increases. The displacement of the top of the medium is given by 
\begin{equation}
    \left. \us \right|_{\zt=1} = -\DPS \frac{\F+\exp(-\F)-1}{\F^2},
    \label{eqn_UsTop_Fluid}
\end{equation}
which is equivalent, for small friction numbers, to
\begin{equation}
    \left. \us \right|_{\zt=1} \sim -\DPS \left(\frac{1}{2}-\frac{\F}{6} \right),
\end{equation}
and, for high friction numbers, to
\begin{equation}
    \left. \us \right|_{\zt=1} \sim - \frac{\DPS}{\F}.
\end{equation}
{\color{ref1}The result of equation \eqref{eqn_UsTop_Fluid} is presented in figure \ref{FIG_Comp}\,(f). This highlights that, for both loading scenarios, the displacement at the top of the medium decreases with increasing $\F$, and that at large $\F$, the two scenarios converge to the same behaviour. At small $\F$, the influence of friction is more pronounced in the piston-driven scenario than in the fluid-driven case, as $\left.\us\right|_{\zt=1}$ decreases more rapidly with $\F$ in the former case.}

\subsection{ Relaxation of the forcing }
\label{subsec_analyticalsolutionDecomp}
If the load on a compressed medium is slowly reduced, it will undergo quasistatic decompression. The initial condition for this decompression is described by equations \eqref{eqn_SigmaPiston} and \eqref{eqn_StressFluidComp}. We denote this initial stress field as $\tilde{\sigma}^\prime_c$ and the corresponding initial imposed effective stress or pressure difference as $\sigs_c$ or $\DPS_c$, respectively. During the decompression phase, the loading varies slowly from $\sigs_c$ or $\DPS_c$ towards 0. As the magnitude of the loading decreases, the tangential frictional forces exerted by the surrounding geometry may retain the material in place (\textit{i.e.}, portions of the material may `stick' rather than `slip') until those tangential forces reach the sliding threshold.

We first examine decompression for the piston-driven configuration. When decompression is initiated, a slipping region immediately nucleates at $\zt=1$ because the material at the top boundary must decompress in order to meet the new (lower) imposed effective stress. This slipping zone grows downward from $\zt=1$ as the applied stress magnitude decreases. Within this slipping zone, the frictional stress attains its maximum value, $\sigf = \mu K \sigp$, corresponding to upward motion ($v_s>0$), and equation \eqref{eqn_ApparitionFriction_simple} becomes
\begin{equation}
    0 = \partial_{\tilde{z}}  \tilde{\sigma}^\prime + \F  \tilde{\sigma}^\prime.
\end{equation}

Integrating this differential equation with the boundary condition $\sigp|_{\zt=1}=-\sigs$ gives the effective stress distribution within the sliding region:
\begin{equation}
    {\sigp} = -{\sigs} \exp \left[{ \F \left( 1-\zt \right)} \right].
\end{equation}

Below this slipping zone, the remainder of the material remains stuck due to friction. In this sticking zone, the effective stress remains equal to $\sigp_c$, as established during the compression phase. Denoting the position of the interface between the slipping and sticking regions as $\zt_\mathrm{slip}$, the complete effective stress field is then
\begin{equation}
    \sigp(\zt) = 
    \begin{cases}
  -\sigs\exp \left[{ \F \left( 1-\zt \right)}\right], & \zt \in[\zt_\mathrm{slip},1] \text{ (slipping region),}\\
  -\sigs_c\exp \left[{ -\F \left( 1-\zt \right)}\right], & \zt \in[0,\zt_\mathrm{slip}] \text{ (sticking region),}
  \label{eqn_sigmaGenFric_Piston}
\end{cases}
\end{equation}
where the value of $\zt_\mathrm{slip}$ front is determined by the requirement that the effective stress profile must be continuous, and is given by
\begin{equation}
\zt_\mathrm{slip} = 1 + \dfrac{1}{2\F} \ln \left( \dfrac{\sigs}{\sigs_c} \right).
\label{Zslip_piston}
\end{equation}
Note that $\zt_\mathrm{slip}>0$ if $|\sigs|>|\sigs_c| e^{-2\F}$. Thus, once the applied stress is reduced beyond a threshold value, (\textit{i.e.} once $|\sigs|\leq|\sigs_c| e^{-2\F}$), the entire medium is slipping.

In the fluid-driven scenario, the effective stress is null at the free boundary ($\zt=1$). Thus, any pressure reduction immediately causes motion and a slipping region nucleates at $\zt=1$ as soon as $\DPS$ begins to drop. In this slipping region, the mechanical equilibrium equation \eqref{eqn_ApparitionFriction_simple} becomes
\begin{equation}
    \DPS = \partial_{\tilde{z}}  \tilde{\sigma}^\prime + \F  \tilde{\sigma}^\prime,
\end{equation}
and, as $\sigp|_{\zt=1}=0$, the effective stress in this regions is equal to
\begin{equation}
\sigp = - \dfrac{\Delta P^\star}{\F} \left\{ \exp \left[\F\left( 1-\zt \right)\right] -1 \right\}.
\label{eqn_StressFluidDecomp}
\end{equation}

Outside of this sliding area, the solid is stuck, and the effective stress remains equal to $\sigp_c$. Therefore, the complete effective stress field is
\begin{equation}
    \sigp(\zt) = 
    \begin{cases}
  - \dfrac{\Delta P^\star}{\F} \left\{ \exp \left[\F\left( 1-\zt \right)\right] - 1 \right\} & \zt \in[\zt_{\mathrm{slip}},1] \text{ (slipping region)}\\
  - \dfrac{\Delta P^\star}{\F} \left\{ 1-\exp \left[-\F\left( 1-\zt \right)\right] \right\} & \zt \in[0,\zt_\mathrm{slip}] \text{ (stuck region)}.
\end{cases}
\end{equation}
The position of the slip front can be obtained as before, by enforcing continuity of the effective stress:
\begin{equation}
    \zt_\mathrm{slip} = 1 + \frac{1}{\F} \ln \left(\frac{\DPS}{\DPS_c} \right).
    \label{Zslip_fluid}
\end{equation}
Similarly to what we described in the piston-driven scenario, the whole medium will slide when $|\DPS|\leq|\DPS_c| e^{-\F}$.

\begin{figure*}
\begin{center}
\includegraphics[height=5.8cm]{"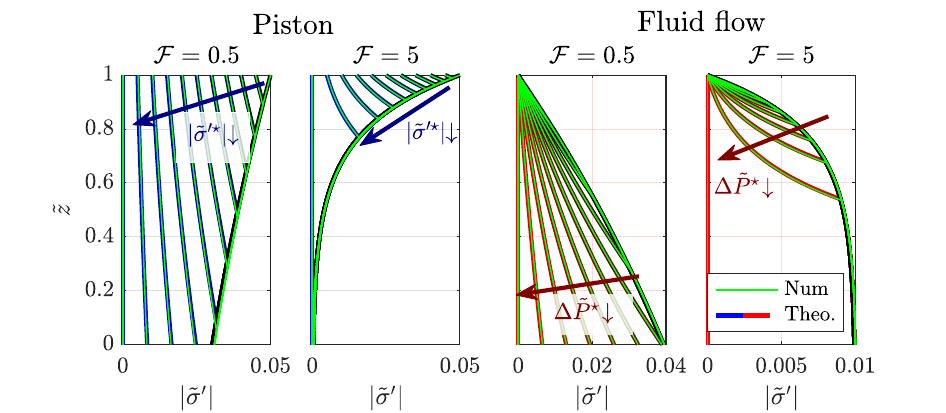"}
\caption{Effective stress as a function of relative position ($\zt$) for a porous medium relaxing from a compressed state (with an applied forcing $|\sigs_c|=|\DPS_c|=0.05 $) toward a fully decompressed state ($\sigs=\DPS=0$) in 11 steps. The loading is imposed by a piston (left half, blue curves) or by a fluid flow (right half, red curves), and the stress field is evaluated from the analytical solutions (continuous coloured curves) and from a full numerical resolution (dotted black). For each forcing, two typical cases are presented: one with relatively low friction ($\F = 0.5$) and one with relatively high friction ($\F = 5$). Arrows show the evolution with decreasing load.}
\label{FIGDecompRealiste}
\end{center}

\end{figure*}
\begin{figure}
\begin{center}
\includegraphics[height=7cm]{"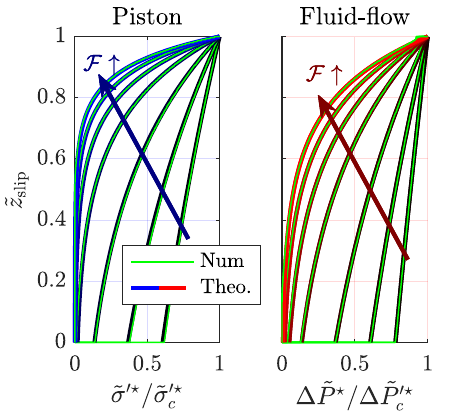"}
\caption{Position of the slip front ($\tilde{z}_\mathrm{slip}$) as a function of the loading intensity, scaled by the intensity of the initial compression ($\sigs_c$, $\DPS_c$), for piston-driven decompression and fluid-driven decompression. In both cases, $\F$ takes the following values: ${0.25, 0.5, 1, 2, 3, 4, 5}$. Theoretical predictions from the analytical model are displayed as solid coloured lines, while numerical results appear as green lines.}
\label{FIG_Ysf}
\end{center}
\end{figure}

We illustrate the above analytical results for piston-driven and flow-driven decompression in figures \ref{FIGDecompRealiste} and \ref{FIG_Ysf}. For each forcing, two typical situations are evaluated: low friction ($\F = 0.5$) and high friction ($\F = 5$). In every situation, the medium has previously been compressed by an imposed stress magnitude of $0.05$. {\color{ref3}In the piston-driven case, this value corresponds to the stress magnitude imposed at the top of the medium; in the fluid-driven case, it represents the stress 
magnitude that would prevail at the bottom in the absence of friction, with the actual stress at the bottom given by equation~\eqref{eqn_StressBottom_fluid}.} The medium is then decompressed gradually, with the imposed stress decreasing from $0.05$ to $0$ in steps of $5 \times 10^{-3}$.  

Recall that these analytical solutions were enabled by several simplifying assumptions. {\color{revisions5}Specifically, we assumed quasistatic decompression --- meaning that the effective stress relaxes gradually without rapid jumps, which may happen during stick-slip motion --- and that slipping zones nucleate at $\zt=1$.} To assess the validity of these assumptions, we also solve the more general model presented in table~\ref{TBL_SystemOfEquations} numerically (see Appendix). These numerical results are compared with the analytical results in figures \ref{FIGDecompRealiste} and \ref{FIG_Ysf}, and closely match our analytical predictions in all scenarios, validating the theoretical approach.


In all cases, and even for relatively small $\F$, a portion of the material remains stuck until the loading is reduced below the corresponding threshold value (see eqs \ref{Zslip_piston} and \ref{Zslip_fluid} above), a signature of the slip front. During this initial phase of the decompression, the stress field is continuous, but its derivative exhibits a jump at the location of the slip front due to the sudden change in material behaviour at this interface. The position of the slip front is presented in figure \ref{FIG_Ysf}, where $\F$ gradually increases. For $\F=5$, a large portion of the medium remains stuck during almost all the decompression phase. Indeed, in this case, some portion of the medium remains stuck until $\sigs / \sigs_c$ falls below $e^{-2\F} = 4.5\times 10^{-5}$ in the piston-driven case and until $\DPS/\DPS_c$ falls below $e^{-\F}=6.7 \times 10^{-3}$ in the fluid-driven case. {\color{revisions5}In any case, the entire material always relaxes to its initial position as long as the stress is reduced below a threshold value. {\color{revisions6}In other words}, there is no residual stress or deformation).}

 \subsection{Apparent behaviour during a compression/decompression cycle}
\begin{figure}
\begin{center}
\includegraphics[height=6cm]{"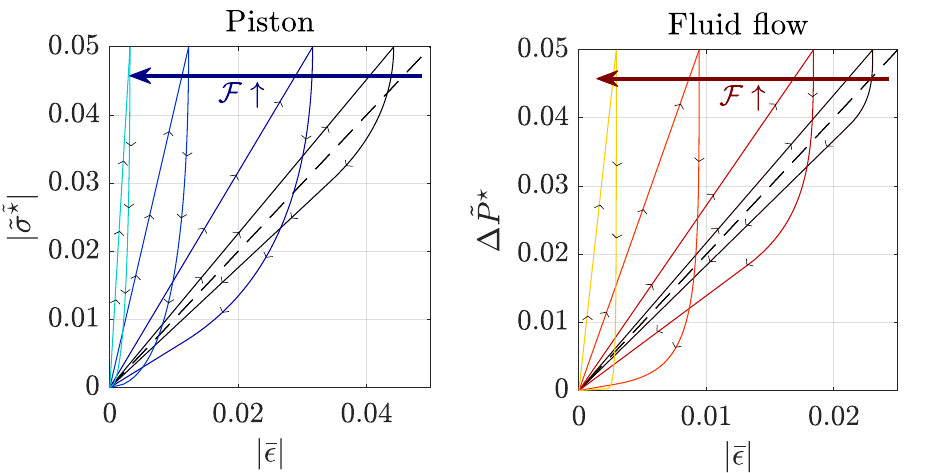"} 
\end{center}
\caption{Nominal macroscopic strain of the porous medium ($\bar{\epsilon}$) as a function of the intensity of the applied loading during a compression/decompression cycle conducted with a piston (left) and with a fluid-flow (right), with arrows following the direction the curves are travelled. Results plotted for the frictionless reference case ($\F=0$, dashed curve) and for $\F \in \{0.25, 1, 4, 16\}$. Note that in the frictionless case, the dashed curve is followed during both compression and decompression. }
\label{FIGHysteresis}
\end{figure}

We now use the results of the previous section to derive the nominal macroscopic strain ($\tilde{u}_s|_{\zt=1}:=\bar{\epsilon}$) during a complete compression/decompression cycle. The results are presented in figure \ref{FIGHysteresis}, showing both hysteresis and apparent stiffening due to friction: a signature of energy dissipated by friction.

\begin{figure}
\begin{center}
\includegraphics[height=6cm]{"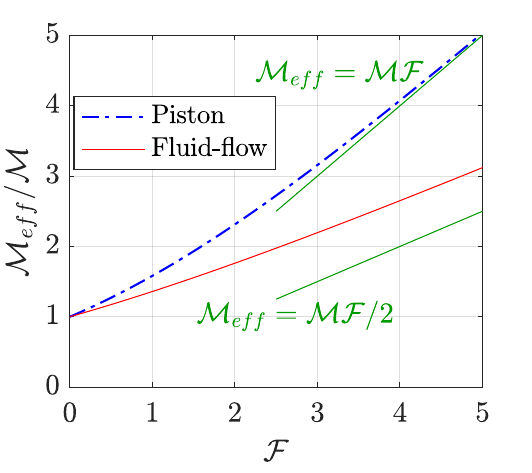"} 
\end{center}
\caption{ Friction leads to an apparent stiffening (\textit{i.e.}, increase in modulus) of the material during compression by a factor $\mathcal{M}_\mathrm{eff}/\mathcal{M}$, as plotted here against the friction number $\F$.}
\label{FIGErrorM}
\end{figure}

During compression, the nominal strain increases linearly with the applied load, but less steeply and to a lower maximum value as friction increases. Therefore, in compression, friction stiffens the apparent response of the material. In the piston-driven case, evaluating equation \eqref{eqn_u_sPiston} at $\zt=1$ shows that the apparent stiffness is equal to
\begin{equation}
    \mathcal{M}_\mathrm{eff}{} = \mathcal{M}\frac{\F}{1-e^{-\F}},
    \label{MeffPiston}
\end{equation}
so that, when $\F \ll 1, ~\mathcal{M}_\mathrm{eff}\sim \mathcal{M}(1+\F/2)$, and when $\F \gg 1$, $\mathcal{M}_\mathrm{eff} \sim\mathcal{M} \F$. In the fluid-driven case, equation \eqref{eqn_usfluide} shows that this apparent stiffness is equal to
\begin{equation}
    \mathcal{M}_\mathrm{eff} = \frac{\mathcal{M}}{2}\frac{\F^2}{e^{-\F}+\F-1}
    \label{MeffFluid}
\end{equation}
so that when $\F \ll 1, 
~\mathcal{M}_\mathrm{eff} \sim \mathcal{M}(1+\F/3)$, and when $\F \gg 1$, $\mathcal{M}_\mathrm{eff} \sim \mathcal{M}\F/2$. {\color{ref1}The results of equations \eqref{MeffPiston} and \eqref{MeffFluid} are presented in figure \ref{FIGErrorM}, together with the two asymptotes $\mathcal{M}_{eff} = \mathcal{M}\F$ and $\mathcal{M}_{eff} = \mathcal{M}\F/2$. This highlights that $\mathcal{M}_{eff}(\F)$ is in fact only weakly nonlinear for both types of forcing.}

\begin{figure}
\begin{center}
\includegraphics[height=5.5cm]{"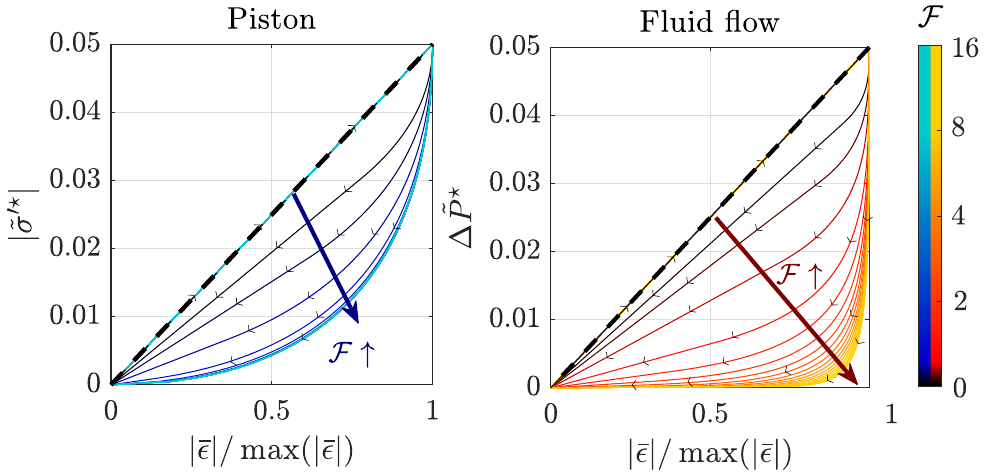"} 
\end{center}
\caption{Nominal strain scaled by the value at the end of compression, as a function of the intensity of the applied load during a compression/decompression cycle conducted with a piston (left) and with a fluid-flow (right), with arrows following the direction the curves are travelled. All the curves follow the diagonal line during compression. The reference frictionless solution ($\F=0$) follows the dashed diagonal, for which compression and decompression paths overlap completely. Results plotted for $\F \in \{0, 0.25, 0.5, 1, 2,3, ..., 16\}$ (with the colour depending on $\F$). }
\label{FIGHysteresisRescaled}
\end{figure}

Figure \ref{FIGHysteresis} also reveals that the piston-driven and fluid-driven cases exhibit fundamentally different behaviour during decompression. This difference can be shown more clearly by scaling the nominal strain by its maximum value (\textit{i.e.} the nominal strain at the end of compression), as presented in figure \ref{FIGHysteresisRescaled}. During piston-driven decompression, the rescaled displacements rapidly collapse to a large-$\F$ asymptotic behaviour: from $\F=4$ to $\F=16$, the curves are almost identical. By contrast, the fluid-driven medium reaches an asymptotic limit at much higher $\F$. To better understand these differences, we next analyse the energy stored, dissipated, and recovered during these hysteresis loops.

\subsection{Energy dissipation}
\label{subsec_energy}
The enclosed area of a compression-decompression cycle is a measure of the amount of energy dissipated by friction during the cycle. We next examine the energetics of this problem in more detail by developing an energy budget that incorporates four key components: elastic energy storage in the solid matrix, work done on the solid matrix by fluid pressure variations, or by the piston, and energy dissipated by friction. To enhance readability, all equations in this section are expressed in their dimensional form (continuing to write $\sigma_{zz}^\prime=\sigma^\prime$, for simplicity).

\renewcommand{\sigp}{{\sigma}^\prime}
\renewcommand{\sigs}{{\sigma}^{\prime \star}}
\renewcommand{\DPS}{\Delta {P}^\star}
\renewcommand{\sigf}{{\sigma}_F^\prime}

During compression, the medium moves from a relaxed state ($u_s = 0, \sigs=0$ or $\DPS=0$) to a compressed state ($u_s = u_{s,c}$ with $\sigs=\sigs_c$ or $\Delta P = \Delta P^{\star}_c$). We denote the stress field inside the medium at this compressed state as $\sigp_c$. The elastic potential energy stored in the solid matrix at the end of the compression is then
\begin{equation}
    E_\mathrm{pot} = \frac{\pi R^2}{2\mathcal{M}}\int_0^L  (\sigp_c)^2 dz.
\end{equation}
For piston-driven compression, the work done for an infinitesimal displacement $\delta u_s |_{z=L}$ is $\sigs \pi R^2  \delta u_s |_{z=L}$. Therefore, the work done on the porous medium by the piston during the full compression is
\begin{equation}
    E_\mathrm{pist}  = \pi R^2 \int_0^{u_{s,c}|_{z=L}} \sigma^{\prime \star } du_s|_{z=L}.
\end{equation}

For fluid-driven compression, the fluid pressure must continuously inject power into the system to move the fluid through the porous medium. During an infinitesimal increment of time $\delta t$, during which the pressure is equal to $\DPS$, the total flux through the porous medium is equal to $q$, the fluid velocity to $v_f$ and the solid velocity to $v_s$. In this case, the total work done on the porous medium by the fluid pressure is equal to
\begin{equation}
    W^+_\mathrm{fluid} = \delta t\DPS q \pi R^2,
\end{equation}
and the portion of this work that is lost to viscous dissipation inside the porous medium is
\begin{equation}
    W^-_\mathrm{fluid} = \delta t \pi R^2 \int_0^L \frac{\eta}{k} \left[ \phi (v_f-v_s)\right]^2 dz.
    \label{eqn_WorkViscous}
\end{equation}
Equation \eqref{eqn_WorkViscous} can be re-written using equations \eqref{eqn_Darcy} and \eqref{eqn_q}:
\begin{equation}
    W^-_\mathrm{fluid} = \delta t \pi R^2 \int_0^L \partial_zP (q -v_s) dz.
\end{equation}
From equation \eqref{eqn_qconst}, $q$ is independent of $z$, and the above becomes
\begin{equation}
    W^-_\mathrm{fluid} = \delta t \pi R^2 \DPS q - \pi R^2 \int_0^L \partial_zP v_s \delta t dz.
\end{equation}
{\color{revisions5}
The net work exerted by the fluid flow to deform the solid matrix is therefore
\begin{equation}
    W^\mathrm{net}_\mathrm{fluid} =W^+_\mathrm{fluid} -W^-_\mathrm{fluid} = \pi R^2 \int_0^L \partial_zP v_s \delta t dz.
\end{equation}
Thus, the net energy imparted to the solid matrix during compression due to fluid pressure is}
\begin{equation}
    E^{net}_{fluid} = \pi R^2 \int_0^{u_{s,c}} \int_0^L \partial_zP du_s dz.
\end{equation}

Separately, the work done by friction on a unit surface $dS$, when the solid matrix moves by an amount $\delta u_s$, is equal to $\sigf dS\delta u_s$. Therefore, the energy dissipated by friction during the compression is equal to
\begin{equation}
    E_{\F} = -\int_0^{u_{s,c}} \int_0^{2\pi} \int_0^L \sigf d u_sRd\theta dz = \pi R^2 \frac{\F}{L} \int_0^{u_s,c} \int_0^L \sigp d u_s dz,
\end{equation}
as during the compression $\sigf= -\F\frac{ R}{L} \sigp $.

Finally, energy conservation during the compression of the porous medium states that
\begin{equation}
E_\mathrm{pot} = E_\mathrm{pist} + E^\mathrm{net}_\mathrm{fluid} - E_{\F},
\end{equation}
which we rewrite as
\begin{equation}
\underset{\gamma^\star}{\underbrace{ \int_0^{u_s^C} \sigs  \delta u_s(L) - \int_0^{u_s^C} \int_0^L  \frac{\DPS}{L} \delta u_s dy }} = 
\underset{\gamma_\mathrm{pot}}{\underbrace{ \frac{1}{2\mathcal{M}}\int_0^L  (\sigma^{\prime}_c)^2 dy }} +
\underset{\gamma_{\F}}{\underbrace{ \frac{\F}{L} \int_0^{u_s^C} \int_0^L \sigp \delta u_s dy }} .
\label{BilanEnergy}
\end{equation}

 \begin{figure}
\begin{center}
\includegraphics[height=6cm]{"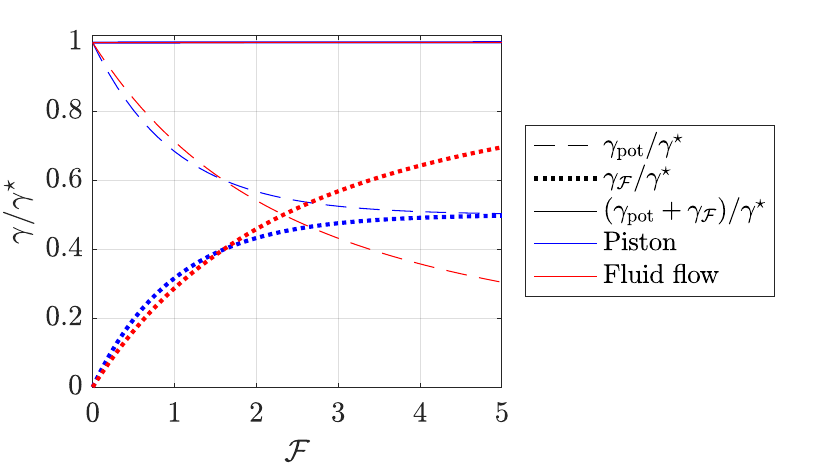"} \\
\end{center}
\caption{Elastic potential energy stored in the solid matrix and energy dissipated due to friction as a function of $\F$, both quantities having been rescaled by the total energy provided to the system, for compression by a piston (blue) and by a fluid flow (red).}
\label{FIGEnergy}
\end{figure}

The left-hand term ($\gamma^\star$) represents the total work done on the solid matrix, either by a piston (where $\sigs>0$ and $\DPS=0$) or by a fluid flow (where $\DPS>0$ and $\sigs=0$). On the right-hand side, the terms $\gamma_\mathrm{pot}$ and $\gamma_\F$ are the elastic potential energy and the energy dissipated by friction, respectively. Note that we have normalised all three energies by the cross-sectional area, $\pi R^2$.

We evaluate each of these terms numerically for each forcing, and at different levels of $\F$, and then normalise $\gamma_\mathrm{pot}$ and $\gamma_\F$ by the total work $\gamma^\star$. The results, as presented in figure \ref{FIGEnergy}, confirm that the amount of energy dissipated by friction increases with $\F$, and the sum of the stored energy and the dissipated energy is always equal to the total work done on the solid (\textit{i.e.}, $\gamma^\star = \gamma_\mathrm{pot} + \gamma_\F$). At low $\F$, both scenarios show similar levels of energy dissipated by friction, so that at $\F=1$, around $30\%$ of the input energy is dissipated by friction for both kinds of loading.

\renewcommand{\sigp}{\sigma^\prime}

For piston-driven compression, the stored elastic potential energy is always larger than the energy dissipated by friction, and both quantities converge to $1/2$ for large $\F$. This relationship is a marker of the coupling between these two energies, which is a consequence of equation \eqref{eqn_EqDiffCompPiston}. Indeed, because of the relationship between the displacement and the strain (equation \ref{DefStrainDepl}), and between the strain and the stress (equation \ref{eqn_StressStrainRelation}), we can rewrite the energy dissipated by friction as
\begin{equation}
    \gamma_\F =  \frac{\F}{L\mathcal{M}} \int_0^{\sigp_c} \int_0^L \sigp (z) \left( \int_0^z  d \sigp (\xi) d\xi  \right)dz,
\end{equation}
where we made a change of variable, writing $u_{s,c}(z) = \int_0^z \sigp(\xi) / \mathcal{M} d\xi$, and therefore $d u_{s,c}(z)=\int_0^zd\sigp(\xi)/\mathcal{M}d\xi$, with $d \sigp$ an infinitesimal increment in effective stress. Using  equation~\eqref{eqn_EqDiffCompPiston}, the inner integral in equation (3.44) can be evaluated: 
\begin{equation}
\begin{split}
\gamma_\F & = \frac{1}{\mathcal{M}} \int_0^{\sigma^{\prime }_c}\int_0^L \left(\sigp(z) \int_0^z\partial_\xi(d\sigp) d\xi \right) dz \\ 
    & = \frac{1}{\mathcal{M}}\int_0^L \int_0^{\sigma^{\prime}_{c}} \sigp(z) \left[  d\sigp(z) - d \sigp(0) \right] dz.
\end{split}
\end{equation}
When $\F$ is large, the effective stress at the bottom of the porous medium is negligible relative to the effective stress at the top (see equations \ref{eqn_StressBottom_piston} and \ref{eqn_StressBottom_fluid}), in which case
\begin{equation}
\gamma_\F \approx   \frac{1}{\mathcal{M}} \int_0^L \frac{(\sigma^{\prime}_c)^2}{2} dy = \gamma_\mathrm{pot}.
\end{equation}
Therefore, {\color{ref3}there exists a the direct coupling between elastic energy storage and frictional energy dissipation}, which explains why for high $\F$ only one-half of the energy given to the system is dissipated by friction.

For fluid-driven compression, equation~\eqref{eqn_EqDiffCompFluid} reveals a more complex scenario, where the frictional stress is not directly linked to the local level of stress because the gradient of fluid pressure adds energy locally (while the piston only adds the energy at the top of the medium).

\begin{figure}
\begin{center}
\includegraphics[height=6cm]{"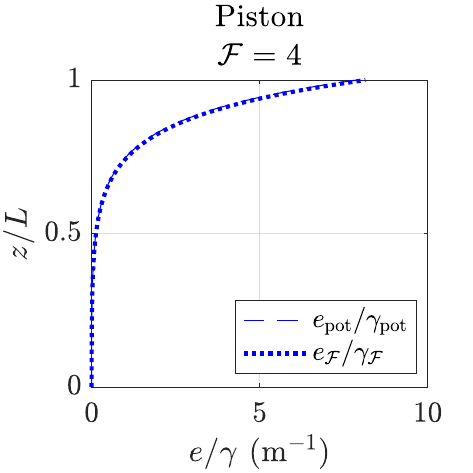"}
\includegraphics[height=6cm]{"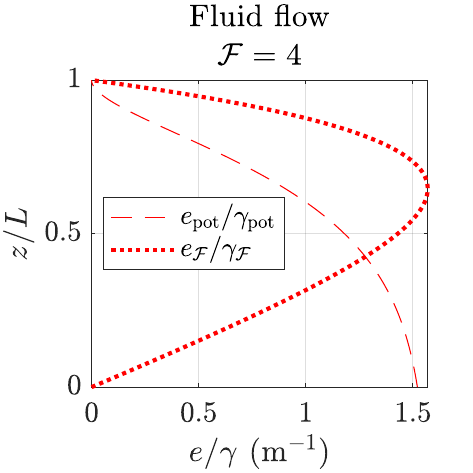"}
\end{center}
\caption{Local energy density during compression as a function of the position, for piston-driven and fluid-driven compression.}
\label{FIGEnergyLocal}
\end{figure}

To illustrate these results, we compute the density of energy that is stored/dissipated at each position $z$. The density of elastic potential energy stored locally is
\begin{equation}
    e_\mathrm{pot} = \frac{1}{2\mathcal{M}} (\sigma^{\prime}_{c})^2,
\end{equation}
while the local energy dissipated by friction is
\begin{equation}
    e_{\F} = \frac{\F}{L}\int_0^{u_{s,c}}\sigp \delta u_s.
\end{equation}
We compute each of these energy densities numerically and normalise the results by the corresponding total energy (\textit{i.e.} we compute $e_\mathrm{pot}/\gamma_\mathrm{pot}$ and $e_\F / \gamma_\F$). The results are presented in figure \ref{FIGEnergyLocal} for a relatively high level of friction ($\F=4$), showing a fundamental difference in energy dissipation and storage inside the porous medium. For piston-driven compression, the energy dissipation and storage are both concentrated at the top of the medium. For fluid-driven compression, the dissipation and the storage of energy are uncoupled: the storage of energy occurs mainly at the bottom of the medium while the dissipation of energy is maximum around the middle (in this case, at $\zt \approx 0.6$). Because of this decoupling, more energy can be dissipated by friction during fluid-driven compression than during piston-driven compression. 

 \begin{figure}
\begin{center}
\includegraphics[height=6cm]{"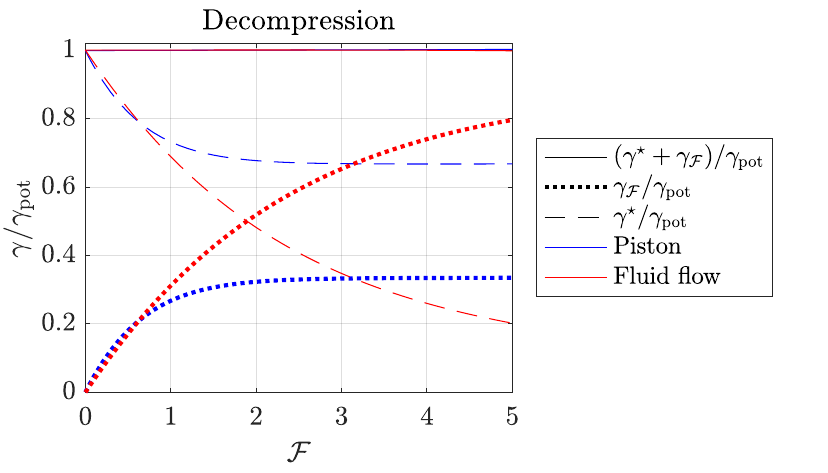"} \\
\end{center}
\caption{Amount of energy dissipated due to friction and released during decompression as a function of $\F$, and rescaled by the total energy stored before the decompression, for a piston (blue) and for a fluid-flow (red).}
\label{FIGEnergyDecomp}
\end{figure}

The same exercise can be conducted during the decompression of the medium. As in section \ref{subsec_analyticalsolutionDecomp}, the decompression is carried out from an initially compressed medium, where the effective stress is equal to $\sigp=\sigp_c$, and pursued until the medium is fully relaxed ($\sigp=0$ everywhere). Friction does work only where the material is sliding, in which case $\sigf = + \frac{\F R}{L} \sigp$. Equation \eqref{BilanEnergy} then becomes
\begin{equation}
    \underset{\gamma_\mathrm{pot}}{\underbrace{ \frac{1}{2\mathcal{M}}\int_0^L  (\sigma^{\prime}_{c})^2 dz }} = 
    \underset{\gamma_{\F}}{\underbrace{ \frac{\F}{L} \int_{u_{s,c}}^0  \int_{z_{slip}}^L \sigp d u_s dz }} +
    \underset{\gamma^\star}{\underbrace{ \int^{0}_{u_{s,c}} \sigs  d u_s(L) - \int^{0}_{u_{s,c}}  \int_0^L  \frac{\DPS}{L} du_s dz }}.
\end{equation}
This energy budget reveals how the medium mobilises its previously stored elastic energy. The stored potential energy ($\gamma_\mathrm{pot}$) acts as an energy reservoir that gets converted into two competing pathways: it is either dissipated through friction or transferred out of the system as useful work (manifested as forces exerted on the boundary or modifications to the fluid flow). This energy partitioning determines the overall efficiency of the energy recovery process.

Following the same computational approach as in compression, we calculate the energy components $\gamma^\star$ and $\gamma_\F$ for different values of $\F$. However, to assess the energy recovery efficiency, we normalise these quantities by the elastic potential energy $\gamma_\mathrm{pot}$ at the onset of decompression, as presented in figure \ref{FIGEnergyDecomp}. This normalisation directly quantifies what fraction of the stored energy is recovered versus irreversibly lost by friction.

The results show again a fundamental change of behaviour between the piston-driven decompression and the fluid-driven decompression for $\F \gg 1$. In the piston-driven case, the energy that is given back to the piston rapidly tends to a value of around $66 \%$ of the stored energy. In the same way, the energy dissipated by friction remains lower than $1/3$ of the initially stored energy, even for high values of $\F$. By contrast, in the fluid-flow case, energy dissipated by friction increases asymptotically toward $100\%$ as $\F$ increases: for $\F=5$, for example, $80\%$ of the initially stored energy has been dissipated by friction, and thus only $20\%$ is returned to the fluid. As before, these contrasting behaviours result from the direct coupling between friction and elastic stress in the piston-driven case. Using the same reasoning as in the compression phase, one can show that, for $\F \gg 1$, $\gamma^\star\approx 2 \gamma_\F$.

Note that our results have been rescaled by the input energy ($\gamma^\star$ during compression, $\gamma_\mathrm{pot} $ during decompression). But, depending on the practical conditions of the test, $\gamma^\star$ can also be sensitive to the friction magnitude. For example, the test may be conducted in a stress-imposed condition, so that the final level of stress at the top of the medium is fixed (say, for example, at $0.05 \mathcal{M}$, as in figure~\ref{FIGHysteresis}). In the presence of friction, the medium will be less easy to move than in the frictionless case (see figure~\ref{FIGHysteresis}). Therefore, in the end, less energy will be given to the system, so that the total amount of stored elastic energy is lower than in the frictionless case. Suppose the test is conducted so that the energy supplied to the system is fixed, as might be the case if the goal were to store a given amount of excess energy for later use. In that case, our results can be directly applied: for high friction coefficients, half of the given energy is stored as potential energy during compression, and during decompression, two-thirds of this stored potential energy can be recovered.

\section{Discussion and conclusions}
\label{sec_Discussion}
This study presents a first theoretical framework for understanding the impact of wall friction on confined poroelastic media. Our analysis establishes three principal contributions: (i) the emergence of the friction number $\mathcal{F}$, which depends on both the friction coefficient and the aspect ratio of the system ($L/R$), as the sole governing parameter for frictional effects (ii) the fundamental distinction between frictional elasticity (piston-driven) and frictional poroelasticity (fluid-driven) regimes, and (iii) the characterization of how friction modifies the apparent mechanical response during quasistatic compression and decompression, including the discovery of slip fronts: spatially heterogeneous decompression patterns that serve as unique signatures of wall friction, distinguishing it from other dissipative mechanisms such as internal rearrangements or plasticity.

{\color{revisions5}For any kind of loading, our framework unifies two classical models: (i) the Janssen model, where the friction exponentially damps both stress and displacements fields toward zero over a (dimensional) characteristic length scale $\lambda_F=L/\F$ \citep{boutreux_propagation_1997,lovisa_tall_2015} and (ii) {\color{ref3}the linear poroelasticity, in which stress obeys a pure diffusion equation in one-dimensional configurations with incompressible constituents \citep[][part 7.3] {cheng_poroelasticity_2016}.} In the presence of friction, that diffusion equation is modified, with the frictional contribution appearing as an advective term scaled by $\F$ (see equation \ref{eqn_AdvDiffStressAdim}), which determines whether the system exhibits frictionless behaviour ($\F \ll 1$) or friction-dominated mechanics ($\F \sim 1$ or larger). {\color{ref2}While a systematic quantitative comparison with experimental data is left for future work, our results agree qualitatively with previous studies. In particular,} the scaling of $\F$ with the aspect ratio clarifies why the aspect ratio has repeatedly been identified as a key control parameter for frictional effects, in both piston-driven and fluid-driven systems \citep[\textit{e.g.}][]{lovisa_consolidation_2012,lutz_frictional_2021}.}

In the piston-driven case, quasistatic loading implies that the fluid has time to drain completely, so there is no pressure gradient and no poroelastic coupling. This scenario reduces to purely elastic mechanics with friction. {\color{revisions5}Our results reproduce the classical Janssen behaviour during compression, and further document the response upon decompression, a regime in which granular materials typically display irreversible deformations due to particle rearrangements \citep{he_dem_2017}. Our solid matrix exhibits fully reversible deformation, owing to its elastic characteristics. Moreover, we demonstrate that at high friction, the dissipation of energy by friction is inherently coupled to the elastic potential energy, so that the energy supplied to the system cannot be entirely dissipated through friction alone.}

The fluid-driven case, in contrast, exhibits full poroelastic coupling, where internal pressure gradients create a \textit{local} forcing, that has profound consequences for both energy dissipation and the mechanical response of the solid matrix. Unlike the piston-driven case, the presence of the fluid flow {\color{ref3}disrupts the direct coupling between elastic energy storage and frictional energy dissipation}. Therefore, during compression, the energy dissipated by friction can substantially exceed the stored elastic potential energy (figure \ref{FIGEnergy}), {\color{revisions5}which may explain why \citet{lutz_frictional_2021} observed different hysteresis shapes when comparing fluid-driven and solid-driven deformations}. Wall friction also alters the poroelastic response: the stress field deviates from its classical linear profile, exponentially approaching a uniform value while displacements transition from quadratic to predominantly linear. This crossover occurs over the same characteristic length $\lambda_\F$ observed in the piston-driven scenario.

For both kinds of forcing, unacknowledged friction introduces systematic measurement errors that could be misattributed to material properties, {\color{revisions5}as already shown by \citet{lu_stress_1998} in the piston-driven case and observed by \citet{lutz_frictional_2021} in the fluid-driven case}. During compression, the apparent elastic modulus increases linearly with $\F$ (figure \ref{FIGErrorM}), such that the classical flow-displacement relationship $u_s|_{z = L} = q\eta L^2/(2k\mathcal{M})$ would lead to permeability being overestimated by similar factors. The dynamic response also reflects these frictional effects. Numerical simulations demonstrate that response times scale as $T_\mathrm{pe, eff} = \eta L^2/(k\mathcal{M}_\mathrm{eff})$, where the friction-enhanced modulus $\mathcal{M}_\mathrm{eff}$ causes {\color{ref3}faster equilibration during loading, but slower equilibration during unloading} (see figure \ref{FIG_Deltaft_F}). Collectively, these errors are particularly insidious because they mimic legitimate material behaviours. Increased apparent stiffness could be misattributed to nonlinear elasticity, while altered permeability might suggest complex microstructural effects. The problem is especially acute in high-aspect-ratio configurations typical of microfluidic devices and filtration columns, where even slippery particles -- such as hydrogels, with friction coefficient $\mu \sim 0.01$ or smaller \citep{cuccia_pore-size_2020} -- can yield large $\mathcal{F}$ values due to the $L/R$ scaling. 

The subtle impacts of friction on the poromechanical behaviour may explain discrepancies reported in previous studies \citep{beavers_flow_1975, parker_steady_1987}. The literature on fluid-driven compression of soft granular media has so far attributed discrepancies between experimental observations and theoretical predictions to rearrangements occurring between the beads \citep{macminn_fluid-driven_2015,hewitt_flow-induced_2016}. Besides, it is well known that the mechanical behaviour of granular assemblies is very different from the behaviour of individual particles \citep{walton_effective_1987,andreotti_granular_2013}, and this mechanical behaviour may depend on subtle properties of the granular packing, such as the number of contacts per particle \citep{luding_discrete_2001}. Therefore, a key question arises: what distinguishes discrepancies caused by the granular nature of the assemblies from those due to friction with the sidewalls, and how can one assess the significance of friction in such experimental studies?

In this context, the emergence of slip fronts during decompression provides a unique fingerprint of wall friction, distinguishing it from other dissipative mechanisms, such as granular rearrangements or plasticity. These fronts, which separate actively deforming regions from zones immobilised by static friction, appear even at low friction numbers and should persist regardless of the material's constitutive behaviour (as this behaviour was not involved in the equations derived in section \ref{subsec_analyticalsolutionDecomp}). In both piston-driven and fluid-driven cycles, the position of such slip fronts evolves logarithmically with the applied force ratio (equations \ref{Zslip_piston} and  \ref{Zslip_fluid}), providing a quantitative diagnostic tool. Experimentalists should therefore be able to distinguish between friction and rearrangement effects by examining residual deformations and monitoring the spatial uniformity of decompression. The presence of stuck regions during partial unloading unambiguously indicates wall friction rather than internal dissipation mechanisms. Apart from these slip fronts, experimental systems could be set up to simultaneously probe fluid pressure, flow rate, and solid deformation, as was done by \cite{lutz_method_2021}. In addition, recently developed photoporomechanical techniques provide a means to observe the local effective stress directly \citep{li_fluid-filled_2022,li_photoporomechanics_2021}. If the solid constitutive law is well characterised (for example, using an independent unconfined piston-driven experiment), an energy budget can be computed, following the methodology we presented in section \ref{subsec_energy}, which could highlight energy dissipation due to friction, a feature discussed by \citet{lutz_frictional_2021}.

Our analytical solutions rely on several simplifying assumptions. We assumed uniform stress across horizontal sections, following classical Janssen-type models, yet it has been well documented that friction introduces radial stress variations in granular silos, {\color{ref2}and that the three-dimensional stress field follows parabolic profiles across the cross section \citep{nedderman_statics_1992,ovarlez_elastic_2005}. As a consequence, using a rigid piston to compress the medium actually imposes a uniform displacement over the top of the medium, which could produce a stress overshoot near the top for finite-sized columns \citep{ovarlez_elastic_2005}. Similarly, the zero-displacement condition at the bottom leads to an abrupt variation of the vertical stress near the lower boundary. These effects are inherently local and do not affect the bulk behaviour captured by our framework, but motivated our choice of a stress-controlled boundary condition at the top. In practice, imposing such a condition requires using a very soft overweight instead of a piston, as mentioned by \citep{ovarlez_elastic_2005}. Besides, because of three dimensional effects,} the stress redirection coefficient $K$ may differ between compression and decompression \citep{nedderman_statics_1992}, which our assumption of a constant $K$ does not capture. Addressing this shortcoming would require distinct friction numbers for compression and decompression, which is analytically tractable, but would require characterising the change in $K$ across different scenarios, which lies beyond the scope of this article. {\color{ref3} Finally, our one-dimensional treatment assumes the slip front to occur at a well-defined altitude. In a three-dimensional framework, the transition between sliding and stuck regions would occur along a 3D portion of the domain. The propagation of the slip front region may then be analogous to the propagation of crack tip in porous media, which produce singularities in the stress field. A full characterisation of these effects would require a dedicated two-dimensional asymptotic analysis and is left for future work.}

Also, we adopted Coulomb friction with equal static and dynamic coefficients, neglecting rate dependence and stick-slip dynamics that are known to occur for soft materials \citep{cuccia_pore-size_2020}. Similarly, we assumed linear elasticity and small strains; finite deformations would require substantially more complex frameworks that couple geometric and material nonlinearities with friction. We also assumed homogeneous, isotropic properties, while real porous materials exhibit spatial variations in permeability and stiffness that could interact with friction in ways that our model cannot capture. Finally, the validity of our analytical solutions is limited to the quasistatic regime. Future work could therefore aim to represent the mechanical behaviour of the frictional medium under cyclic loading at a range of speeds, which is not analytically tractable. To study these complex situations, the numerical solution presented in the Supplementary Material should be used. A key question remains whether slip fronts can emerge in these complex situations. In particular, in fast-motion scenarios, we expect the fluid flow to play a role that has not been explored in this study. Future work could also incorporate the elasto-plastic mechanical behaviour of the solid matrix to account for rearrangements, as \textit{both} rearrangements and friction may be present in some experimental situations.

\backsection[Acknowledgements]{We thank M. Delarue, P. Joseph and P. Duru for several fruitful discussions. We thank L. Morrow for useful discussions on frictional poroelasticity and numerical simulations.}

\backsection[Funding]{This work was supported by MEGEP doctoral school and CNRS PEPS BiBou project. It was also supported by the European Research Council (ERC) under the European Union’s Horizon 2020 Programme (Grant No. 805469) and by the UK Engineering and Physical Sciences Research Council (EPSRC) (Grant No. EP/S034587/1).}

\backsection[Declaration of interests]{The authors report no conflict of interest.}

\backsection[Data availability statement]{The simulation datasets used in this paper are available from the corresponding author upon request.}

\backsection[Author ORCIDs]{T. Desclaux, https://orcid.org/0000-0002-1822-4593; C. Cuttle, https://orcid.org/0000-0002-1852-5599; C.W. MacMinn, https://orcid.org/0000-0002-8280-0743; O. Liot, https://orcid.org/0000-0001-5192-0120}

\section*{Appendix: Numerical solution}
The system of equations presented in table \ref{TBL_SystemOfEquations} is mathematically closed but not analytically solvable in the general case. Therefore, numerical solution of these equations was conducted to check the validity of the analytical solutions presented in section \ref{subsec_analyticalsolution}. In particular, the numerical solution is designed to solve the complete model (equation \ref{eqn_CoulombIntermediate}), allowing our analytical solutions, corresponding to idealised cases, to be evaluated against such realistic scenarios.

\subsubsection*{Solving the friction term}
Coulomb's law was written in equation \ref{eqn_CoulombIntermediate}. In the case where $v_s = 0$, this equation does not directly give the value of $\sigma^\prime_\F$: it only bounds its magnitude. But, it does tell us that $v_s = 0$. 

Besides, the definition of $q$ \eqref{eqn_q} and the Darcy law \eqref{eqn_Darcy}, lead to
\begin{equation}
v_s = q + \dfrac{k}{\eta} \partial_z P.
\label{eqn_vs_DP}
\end{equation}
So that the equation ``$v_s = 0$'' may be re-written, by using the mechanical equilibrium \eqref{eqn_ApparitionFriction_simple}
\begin{equation}
\sigma_F^\prime = -\dfrac{R\eta}{2k} q - \dfrac{R}{2} \partial_z \sigp.
\label{eqn_sigmaF_NOSPEED}
\end{equation}
And that equation is valid if and only if $\left| \sigma_F^\prime \right| \leq \mu K \left| \sigp \right|$. That inequality is itself equivalent to $\left| \frac{R\eta}{2k} q + \frac{R}{2} \partial_z \sigp \right|\,\leq\,\mu\,K \left| \sigp \right| $.

In the end, Coulomb's law can be written
\begin{equation}
\begin{cases}
\sigma_F^\prime = \mathrm{sgn}(v_s) \mu K \sigp & \iff  \left| v_s \right| > 0, \\

\sigma_F^\prime = -\dfrac{R\eta}{2k} q - \dfrac{R}{2} \partial_z \sigp 
& \iff
\left| \dfrac{R\eta}{2k} q + \dfrac{R}{2} \partial_z \sigp \right| \leq \mu K \left| \sigp \right| .
\end{cases}
\label{eqn_CoulombFinal}
\end{equation}

Therefore, in the numerical code, the proper expression of $\sigma'_F$ is derived by testing -- everywhere in the medium and at every time step -- if the following inequality is verified
\begin{equation}
\left| q + \dfrac{k}{\eta} \partial_z \sigp \right| \leq \dfrac{2 k \mu K}{R \eta} \left| \sigp \right|.
\label{eqn_Condition}
\end{equation}
If that condition is true, then
\begin{equation}
\sigma_F^\prime = -\dfrac{R\eta}{2k} q - \dfrac{R}{2} \partial_z \sigp, \mathrm{and, }~ v_s = 0, 
\end{equation}
else, equation \ref{eqn_CoulombFinal} implies that
\begin{equation}
\left| \sigma_F^\prime \right| = \mu K \left| \sigp \right|,
\end{equation}
and therefore, equations \eqref{eqn_vs_DP} and \eqref{eqn_ApparitionFriction_simple} state that
\begin{equation}
\mathrm{sgn}(v_s) = \mathrm{sgn} \left(q + \dfrac{k}{\eta} \partial_z \sigp + \dfrac{2 k \mu K}{R \eta}  \sigp \right) = \mathrm{sgn} \left(q + \dfrac{k}{\eta} \partial_z \sigp \right),
\end{equation}
and, because the negation of equation \eqref{eqn_Condition} is verified
\begin{equation}
\mathrm{sgn}(v_s) = \mathrm{sgn} \left(q + \dfrac{k}{\eta} \partial_z \sigp \right).
\end{equation}
In the end
\begin{equation}
\sigma_F^\prime  = \mathrm{sgn} \left(q + \dfrac{k}{\eta} \partial_z \sigp \right) \mu K \sigp ,
\end{equation}
and
\begin{equation}
v_s = q + \dfrac{k}{\eta} \partial_z \sigp + \mathrm{sgn}(q + \dfrac{k}{\eta} \partial_z \sigp ) \dfrac{2k}{\eta R} \mu K \sigp.
\end{equation}

\subsubsection*{Practical implementation of the code}
The code employs an iterative scheme to handle the non-linear nature of Coulomb friction, which introduces a stick-slip discontinuity in the governing equations. At each time step, an inner iterative loop determines the mechanical equilibrium by identifying which regions of the medium are sliding (where the frictional stress reaches $\mu K \sigma'$) versus stuck (where $v_s = 0$). This iteration is essential because the friction force depends on the stress field, which itself depends on whether regions are sliding or stuck — creating an implicit problem that cannot be solved directly. The algorithm tests each spatial location to determine if the driving force exceeds the maximum static friction, updating the velocity field and stress distribution until convergence is achieved (typically when stress changes fall below $10^{-6}$ and boundary conditions are satisfied). Once mechanical equilibrium is established, the code advances the porosity field using the continuity equation, with boundary conditions enforced to maintain zero displacement at the fixed base.

\subsubsection*{Discretisation}
{\color{revisions5}We transform the system of equations from the Eulerian frame ($z,t$) to a normalized moving frame $(\xi, \tau)$ where $\xi = z/L$ and $\tau = t$, which tracks the solid matrix boundaries (as $L$ evolves in time), and maps the deforming domain to the fixed interval $[0,1]$. This change of variables simplifies boundary condition handling and enables future extension to more general nonlinear poroelastic formulations beyond the linear theory presented here.}

The problem is discretised, following the finite differences method. First, the $\xi$ coordinate is discretised on $N$ points ($\xi_i$ with $i \in [1, N]$), with a constant space step ($\xi_{i+1}-\xi_{i} = 1/N, \forall i \in [1,N]$). Away from the boundaries, the spatial derivatives are approximated by a central-difference scheme. This spatial derivative is precise at the second order, but it cannot be computed at the top (and bottom) boundaries. At these points, the spatial derivatives are computed by a backwards difference scheme (at $\xi=1$) and a forward difference scheme (at $\xi=0$). These schemes are precise at the first order, but are practically easy to implement. The time is also discredited explicitely by the forward Euler method.

\subsubsection*{Discussion of the discretisation method}
For large time steps, the time integration method is not stable (the numerical solution may diverge from the theoretical solution) or accurate (significant errors may be produced) \citep{ferziger_computational_2002}. In the same way, the spatial discretisation method is only accurate at the first order in $\xi_{i+1}-\xi_{i}$. Preliminary convergence tests (not shown) required a small temporal resolution ($\Delta t = 2.5 \times 10^{-6} ~T_{pe}$) and spatial discretisation ($N = 300$) to achieve convergent solutions. Even with such a small resolution, solving a compression/decompression cycle requires $\approx 1$ hour on a desktop computer. This implementation provides reliable and accurate results that effectively validate our analytical framework, the primary goal of the numerical simulations.

\subsubsection*{How slow is quasistatic forcing ?}
The analytical solutions presented in section \ref{subsec_analyticalsolution} assume quasistatic conditions, but the validity of this assumption requires clarification: how slowly must the loading be applied to achieve this regime? To address this question, we examined the step response of the porous medium through numerical simulations. The applied forcing ($\tilde{\sigma}^{\prime\star}$ or $\Delta \tilde{P}^\star$) was ramped from 0 to $0.05$ {\color{ref3}over a time equal to the poroelastic time scale $T_\mathrm{pe}$ (defined equation \ref{eqn_Tpe}). The forcing was then} held constant for $100~T_\mathrm{pe}$, then decreased back to zero over a time $T_\mathrm{pe}$, and maintained at zero for an additional $100~T_\mathrm{pe}$. This loading protocol was applied for both the reference frictionless case and for $\F \in \{0, 0.25, 0.5, 1, 2,3,4,5\}$.

\begin{figure*}
\begin{center}
\includegraphics[width=\linewidth]{"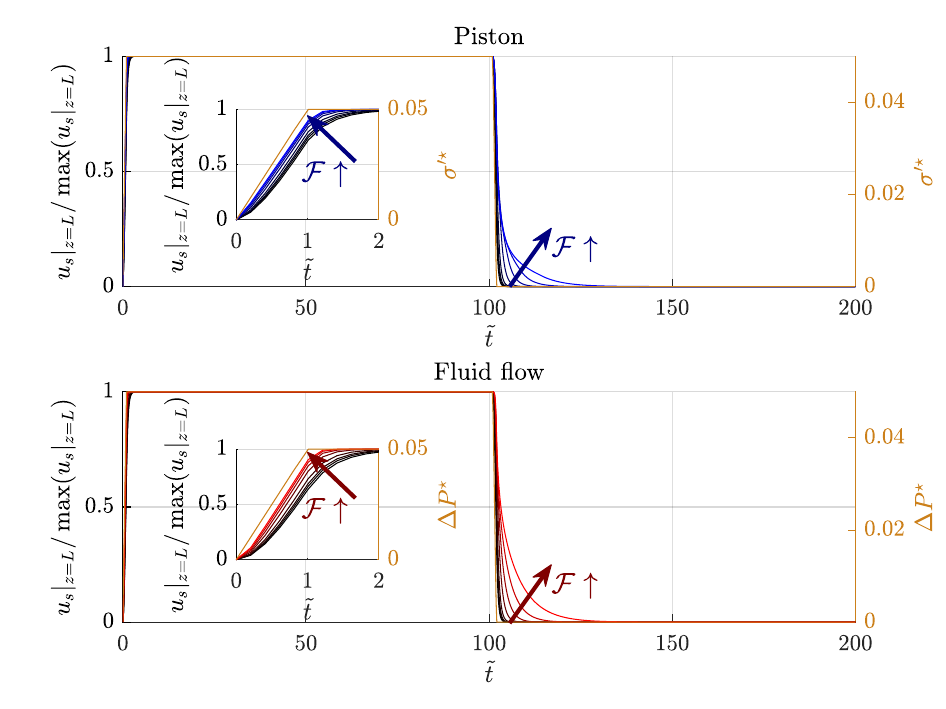"}
\caption{Time evolution of the displacement of the top of the medium, normalised by the maximum displacement, observed at the end of the compression, forcing the medium with a piston (top) and fluid-flow (bottom). The time evolution of the applied forcing is represented in orange (right axis). The inset is a zoom of the early times. 
}
\label{FIG_Deltaft_F}
\end{center}
\end{figure*}

Figure \ref{FIG_Deltaft_F} shows the temporal evolution of the top boundary displacement ($|\tilde{u}_s|_{z=L}$). The results show that during the compression phase, the response time of the porous medium decreases as friction increases, while during decompression, it increases as friction increases. In any case, the porous medium reaches steady state within $25~T_\mathrm{pe}$, following each loading change. 

Therefore, for comparison with the quasistatic solutions, it was decided that the medium would be compressed following a ramp from 0 to $0.05$  over $100~T_\mathrm{pe}$, then remain constant for $100~T_\mathrm{pe}$, then decrease back to $0$ in $100~T_\mathrm{pe}$, before remaining zero during $100~T_\mathrm{pe}$.

\subsubsection*{Comparison with the analytical results}
\begin{figure*}
\begin{center}
\includegraphics[width=\linewidth]{"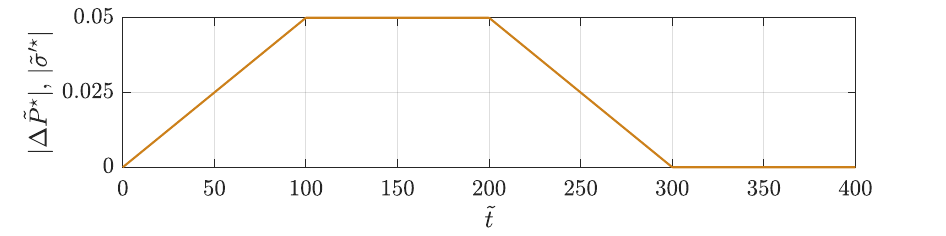"}
\caption{Time evolution of the intensity of the forcing, in the ``quasistatic'' numerical simulation.}
\label{FIG_ForcingNum}
\end{center}
\end{figure*}

Both piston-driven and fluid-driven configurations were examined, with the applied stress varying as explained in the previous part (see figure \ref{FIG_ForcingNum}). For each forcing intensity presented in figure \ref{FIGDecompRealiste}, the stress field is extracted from simulation results. Also, the position of slip fronts is tracked: sliding regions are identified as those where the porosity field differs from the one obtained at the end of the compression by more than $5\times10^{-5}$. This threshold effectively distinguishes actively deforming regions from those immobilised by static friction.

\bibliographystyle{jfm}
\bibliography{Biblio_Terence}
\newpage

\end{document}